\documentclass[review]{elsarticle}

\usepackage{hyperref}
\usepackage{amsmath}
\usepackage{multirow}

\usepackage{booktabs}
\usepackage[table,xcdraw]{xcolor}

\newcommand\myeq{\mathrel{\overset{\makebox[0pt]{\mbox{\normalfont\tiny\sffamily def}}}{=}}}
\newcommand{\me}{\mathrm{e}}

\journal{Computerized Medical Imaging and Graphics}

\usepackage{natbib}
\setcitestyle{authoryear,round}
\begin{document}

\begin{frontmatter}


\title{Computer Aided Detection of Deep Inferior Epigastric Perforators in Computed Tomography Angiography scans}

\author[mymainaddress,mysecondaryaddress]{Ricardo J. Ara{\'u}jo\corref{mycorrespondingauthor}}
\cortext[mycorrespondingauthor]{Corresponding author}
\ead{ricardo.j.araujo@inesctec.pt}

\author[mythirdaddress]{Vera~Garrido}
\author[myfourthaddress]{Catarina~A.~Bara{\c c}as}
\author[mythirdaddress]{Maria~A.~Vasconcelos}
\author[mythirdaddress]{Carlos~Mavioso}
\author[mythirdaddress]{Jo{\~a}o~C.~Anacleto}
\author[mythirdaddress,myfifthaddress]{Maria~J.~Cardoso}
\author[mymainaddress,mysecondaryaddress]{H{\'e}lder~P.~Oliveira}



\address[mymainaddress]{INESC TEC, Campus da Faculdade de Engenharia da Universidade do Porto, Rua Dr. Roberto Frias, 4200-465 Porto, Portugal}
\address[mysecondaryaddress]{Faculdade de Ci{\^e}ncias da Universidade do Porto, Rua do Campo Alegre, 4169-007 Porto, Portugal}
\address[mythirdaddress]{Breast Unit, Champalimaud Clinical Center, Avenida de Bras{\'i}lia, 1400-038 Lisboa, Portugal}
\address[myfourthaddress]{Hospital Pedro Hispano, Rua Dr. Eduardo Torres, 4464-513 Senhora da Hora, Portugal}
\address[myfifthaddress]{Nova Medical School, Campo M{\'a}rtires da P{\'a}tria 130, 1169-056 Lisboa, Portugal}

\begin{abstract}
The deep inferior epigastric artery perforator (DIEAP) flap is the most common free flap used for breast reconstruction after a mastectomy. It makes use of the skin and fat of the lower abdomen to build a new breast mound either at the same time of the mastectomy or in a second surgery. This operation requires preoperative imaging studies to evaluate the branches - the perforators - that irrigate the tissue that will be used to reconstruct the breast mound. These branches will support tissue viability after the microsurgical ligation of the inferior epigastric vessels to the receptor vessels in the thorax. Usually through a Computed Tomography Angiography (CTA), each perforator, diameter and direction is manually identified by the imaging team, who will subsequently draw a map for the identification of the best vascular support for the reconstruction. 

In the current work we propose a semi-automatic methodology that aims at reducing the time and subjectivity inherent to the manual annotation. In 21 CTAs from patients proposed for breast reconstruction with DIEAP flaps, the subcutaneous region of each perforator was extracted, by means of a tracking procedure, whereas the intramuscular portion was detected through a minimum cost approach. Both were subsequently compared with the radiologist manual annotation.

Results showed that the semi-automatic procedure was able to correctly detect the course of the DIEAPs with a minimum error (average error of 0.64 mm and 0.50 mm regarding the extraction of subcutaneous and intramuscular paths, respectively). The objective methodology is a promising tool in the automatic detection of perforators in CTA and can contribute to spare human resources and reduce subjectivity in the aforementioned task.
\end{abstract}

\begin{keyword}
breast reconstruction\sep deep inferior epigastric artery perforator flap\sep computed tomography angiography\sep vessel detection 
\end{keyword}

\end{frontmatter}

\section{Introduction}

Worldwide, there will be about 2.1 million newly diagnosed female breast cancer cases in 2018, accounting for almost 1 in 4 cancer cases among women. The disease is the most frequently diagnosed cancer in the vast majority of the countries and is also the leading cause of cancer death in over 100 countries~\citep{Bray:2018}. Women who were diagnosed with breast cancer have higher chance of suffering from anxiety and depression resulting from the fear of recurrence, body image disruption, sexual dysfunction and mortality concerns~\citep{Hewitt:2004}. Although breast conservative methods have recently shown a survival rate superior to mastectomy, especially in early breast cancer cases~\citep{Gentilini:2017}, the latter is still a highly recurrent procedure and has even been increasing in some institutions~\citep{Dragun:2012,Mahmood:2013}. Mastectomy is conducted in cases where the relation between the size of the resected breast and the global volume of the gland is too large to allow a conservative procedure, in cases where radiotherapy is contra-indicated, and also when the patient does not desire breast conservation.

\begin{figure}[t!]
	\begin{center}
		\includegraphics[width=0.7\columnwidth]{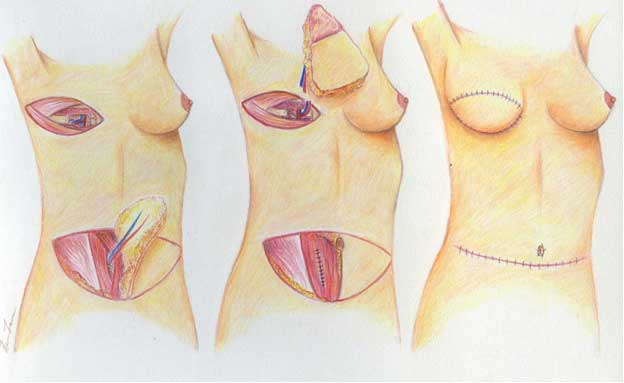}
	\end{center}
	\caption{Representation of a DIEAP flap procedure.}
	\label{fig:image_procedure}
\end{figure}

\begin{figure}[t!]
	\begin{center}
		\includegraphics[width=0.8\columnwidth]{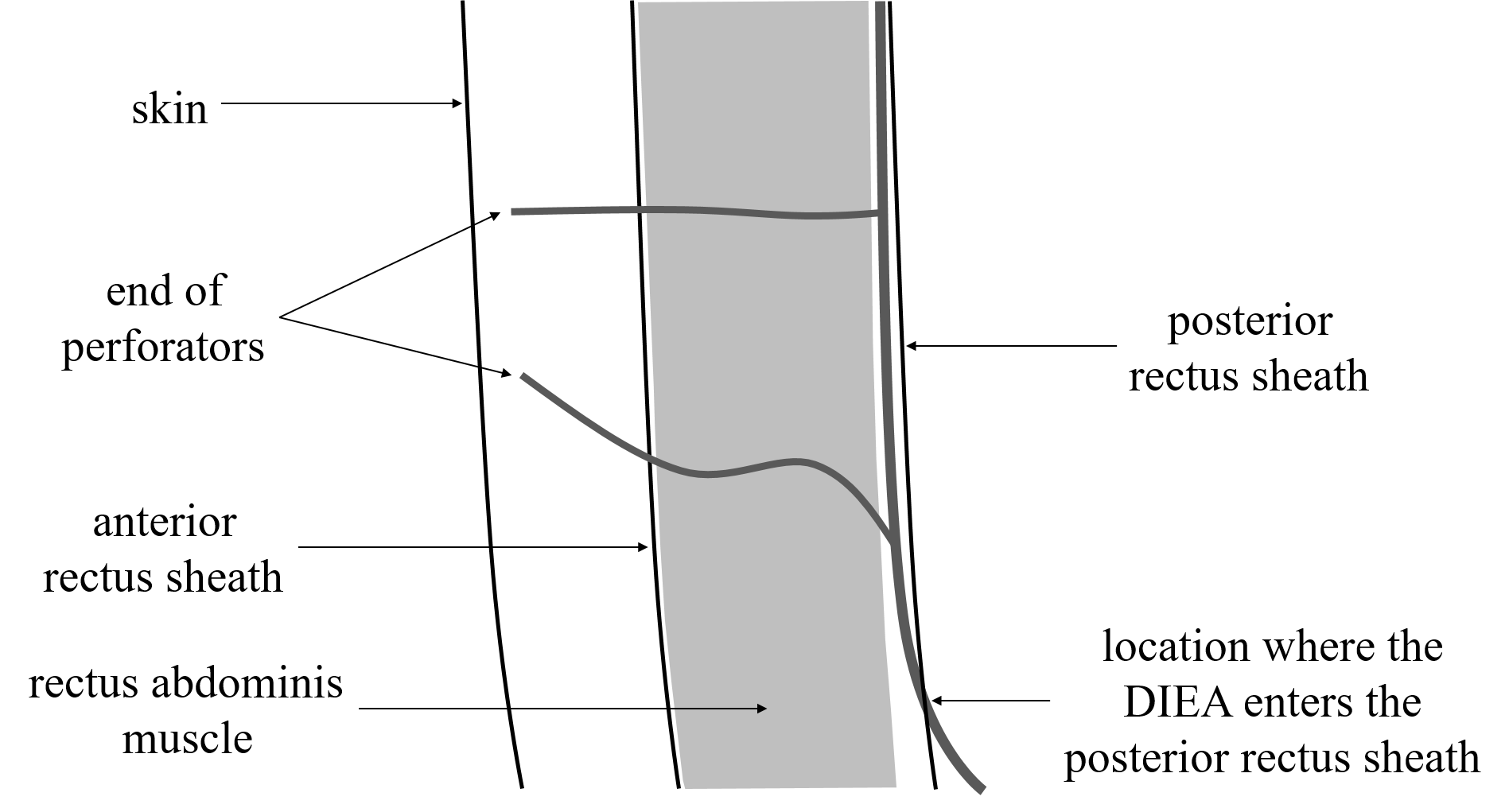}
	\end{center}
	\caption{Sagittal representation of the anatomy of the anterior portion of the abdominal wall, between the pelvic and umbilicus regions.}
	\label{fig:sag_anatomy}
\end{figure}

Reconstruction methods allow to recreate the breast mound, improving the way women feel about themselves and their image after their breast(s) was(were) removed. There are different techniques for breast reconstruction but basically two major groups can be defined. Reconstruction with implants and reconstruction with autologous tissues. The deep inferior epigastric artery perforator (DIEAP) flap has become the state-of-art technique for autologous tissue based breast reconstruction~\citep{Cina:2010}. It makes use of the skin and fat of the lower abdomen to build a new breast either at the same time of the mastectomy (immediate reconstruction) or in a second surgery after the initial procedure (delayed reconstruction). The transposition of the lower abdominal skin and fat is free of any attachment to the end anatomic structures of the donor site - the abdomen. Micro-surgical connections are done at the recipient site between the vessels of the transposed skin and fat, and the vessels of the thorax, where the new breast will replace the void left by the mastectomy (see Figure~\ref{fig:image_procedure}). A scheme with the abdominal anatomy of interest for conducting a DIEAP flap is shown in Figure~\ref{fig:sag_anatomy}.

Before a DIEAP flap, preoperative imaging studies are performed to evaluate the branches - perforators - of the deep inferior epigastric artery (DIEA), which are the vessels responsible for the vascularization of the tissue that will be used in the reconstruction of the breast mound. The viability of the new breast is related to several features of the included perforators~\citep{Phillips:2008}. Through Magnetic Resonance Imaging (MRI) or Computed Tomography Angiography (CTA), the perforators are manually identified and characterized by a radiologist. This manual map guides the surgeon during the reconstructive procedure. The before mentioned task of identifying and characterizing the 3D course of the perforators is subjective and time consuming. As a result, incoherencies between the preoperative studies and the surgical findings often exist, and can lead to the need of modifying the strategy intra-operatively. For that reason, Computer Aided Detection algorithms may play an important role in supporting the activity of radiologists who are responsible for the preoperative study, reducing the subjectivity and time involved in the process. Moreover, more precise and complete descriptions may be achieved, as it is common to find segments of the perforators whose signal is almost absent, rendering the manual analysis very difficult. Other authors have developed plugins for medical software trying to render the manual analysis of the perforators faster \citep{Lange:2017} and used virtual tools to facilitate the communication of the manual findings with the surgical team \citep{Gomezcia:2009}. In this paper, we propose a methodology for the semi-automatic extraction of the course of the perforators in CTA scans, by means of a vessel centerline extraction technique particularly designed to address the challenges involved in detection of these small vessels. Our goal is to reduce the time involved in the pre-operative planning of DIEAP flaps and increase the overall objectiveness.

\subsection{Related work}

To the best of our knowledge, there is no algorithm in the literature focusing the segmentation of the deep inferior epigastric perforators. Still, vessel segmentation algorithms usually follow common principles and assumptions that stand true for different types of vessels. Even though it does not include the most recent work, a thorough description of the main types of approaches regarding 3D vessel segmentation is available~\citep{Lesage:2009}. Supervised learning techniques have seen very limited applicability in this scenario, as obtaining annotations of the vasculature in 3D scans is a very expensive and difficult task. This goes against the trend that can be currently observed in many domains of Computer Vision and Medical Image Processing, such as the case of vessel segmentation in 2D retinal fundus images~\citep{Liskowski:2016}, where databases with annotations are available \citep{Staal:2004,Budai:2013}.

A myriad of unsupervised methods have been proposed to address the vessel segmentation problem in different 3D scenarios, such as angiograms~\citep{Cruz:2016} and CTA scans \citep{Metz:2009,Friman:2010} of the coronary arteries, CT scans of the lung \citep{Zhai:2016}, and MRA \citep{Elbaz:2012} and CTA \citep{Babin:2013} scans of the brain. The characteristics of the proposed methodologies is often related with the complexity of the vessel network in each case. Centerline tracking approaches are more commonly applied to networks where only a small number of vessels needs to be extracted, as is the case of coronary vessels. More intricate networks, such as the lung and brain vessel trees, are usually segmented by means of more global frameworks, like level sets and graph cuts~\citep{Zhai:2016}. The DIEAPs resemble the case of coronary vessels, in what concerns the complexity of the vessel network. Additionally, the extraction of characteristics from these networks is relevant in a clinical point of view, making centerline based approaches very interesting, as they allow to easily extract several local features of each vessel segment as the vessel is traversed.

Vessel centerline extraction is usually achieved via tracking methodologies or minimum cost path approaches. Centerline tracking schemes iteratively traverse the vessel, allowing for its simultaneous extraction and local characterization. These schemes differ from each other according to the strategy they employ to locally estimate the vessel direction. Recent works track vessels by iteratively finding the best approximation for their cross section~\citep{Kumar:2015}. However, relying in such information is not adequate in a scenario where vessel diameter spawns across very few voxels, thus leading to poor definition of the vessel boundary, as is the case of many epigastric perforating vessels in CTA. Other approaches use the complete 3D local intensity information to track tubular-like patterns, by fitting priorly designed filters~\citep{Friman:2010}, being more robust when tracking small vessel signals. 
Minimum cost path schemes use geometrical features in order to find voxels which more likely belong to vessel patterns~\citep{Metz:2009}. After deriving the costs from those features, the vessel centerline is extracted at a single step. Nonetheless, it is straightforward to have a second scheme for lumen characterization. It is also possible to enhance vessels by means of geometrical-based enhancement followed by thresholding and skeletonization to obtain vessel centerlines~\citep{Yang:2012}. However, such approach is easily affected by local noise, and setting a threshold that works well in every case is not a trivial task. There is an interest in completely automatic centerline extraction approaches~\citep{Oliveira:2016}, but this may have more disadvantages than advantages in scenarios where structured noise may be easily mistaken as a vessel, thus inducing false positives. Moreover, semi-automatic approaches do not constitute a significant burden when only few vessel segments need to be studied, especially when compared with the high cost of manual analysis, as is the case of the perforators targeted in this study.

\subsection{Contributions and structure of the paper}

The contributions of the current work are the following:

\begin{itemize}
	\item the proposal of a methodology to extract in an accurate and objective manner the complete course of the DIEAPs;
	\item the proposal of a tracking procedure for extracting the subcutaneous portion of the perforators based on the local gradient field of a vessel enhanced volume;
	\item the validation of an A* based path search using costs derived from a vessel enhanced volume to extract the intramuscular course of the perforators.
\end{itemize}

The structure of the paper is as follows: section~\ref{sec:methods} describes the database we have available and the methods used to accomplish the contributions of this work, section~\ref{sec:results} contains a view over the results obtained and their discussion and, finally, section~\ref{sec:conclusion} details the conclusions we take on the topics we have addressed.

\section{Materials and Methods} \label{sec:methods}

In this paper, we aim at extracting both subcutaneous and intramuscular courses of the perforators, since they have impact in the pre-operative planning of DIEAP flaps. There is a higher contrast between the vessels and the background in their subcutaneous portion, since the rectus abdominis muscle also responds significantly to the CTA image acquisition. This motivated us to pursue different strategies when addressing the extraction of the subcutaneous and intramuscular courses of the perforators. Anatomically, those regions are separated by the anterior fascia of the rectus abdominis muscle. Unfortunately, that tissue layer is not distinguishable in CTA scans. Nonetheless, it can be approximated as the edge that exists between the subcutaneous region and the muscle. In a previous work~\citep{Araujo:2017}, we have proposed a methodology to segment this layer, thus enabling the employment of a divide-and-conquer strategy to extract the complete course of the perforators. In this section, we start by describing the dataset that we have available and the anatomical region of the volumes we consider for analysis. Afterwards, we propose a novel centerline tracking approach for the extraction of the subcutaneous portion of a perforator, from the initial manually given point to the fascia segmentation. Finally, we describe a minimum cost path based approach to extract the intramuscular path, which is a more complex task.

\subsection{Dataset}

\begin{figure*}[t!]
	\begin{center}
		\begin{minipage}{\textwidth}
			\begin{center}
				\includegraphics[height=40mm,width=40mm]{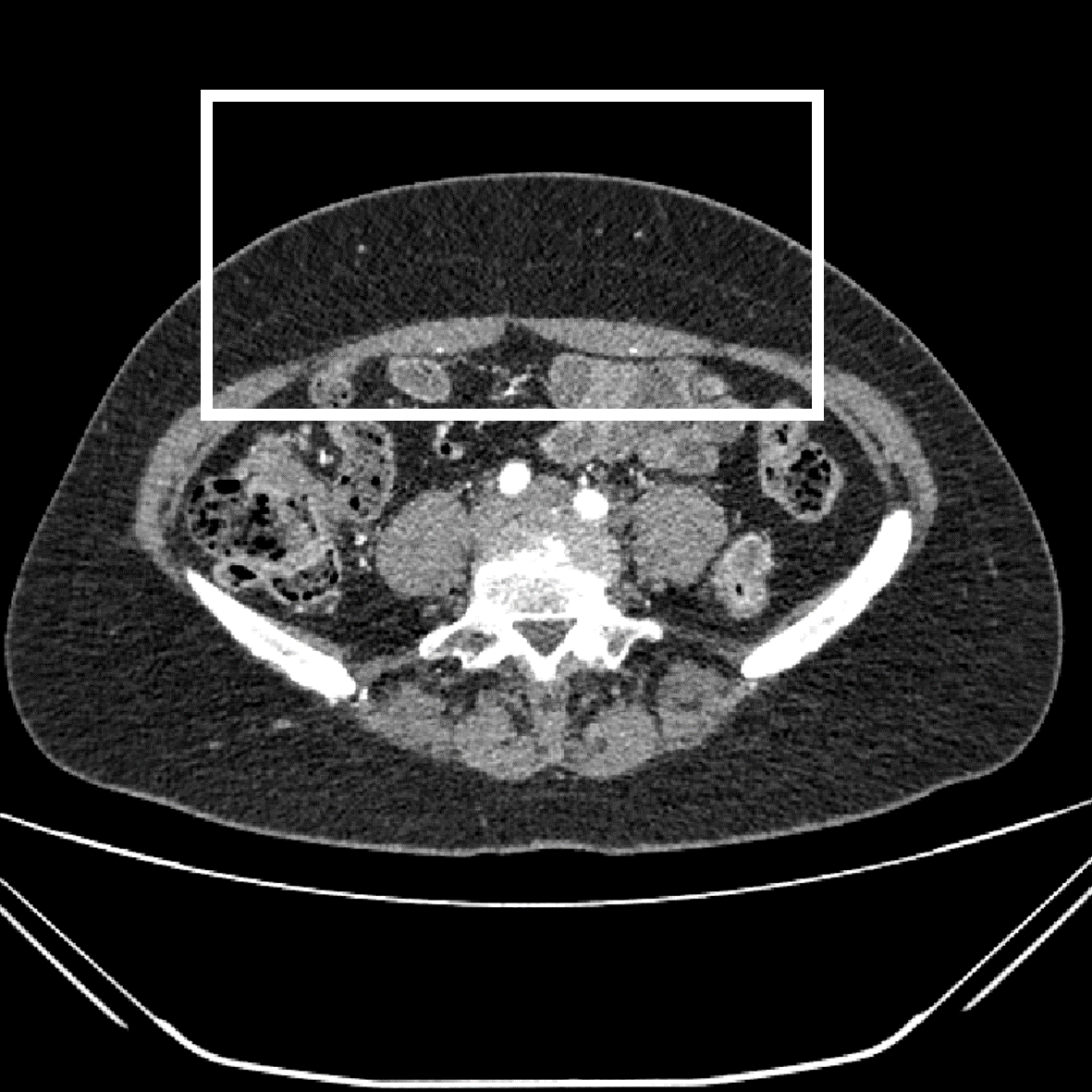}
				\includegraphics[height=40mm,width=70mm]{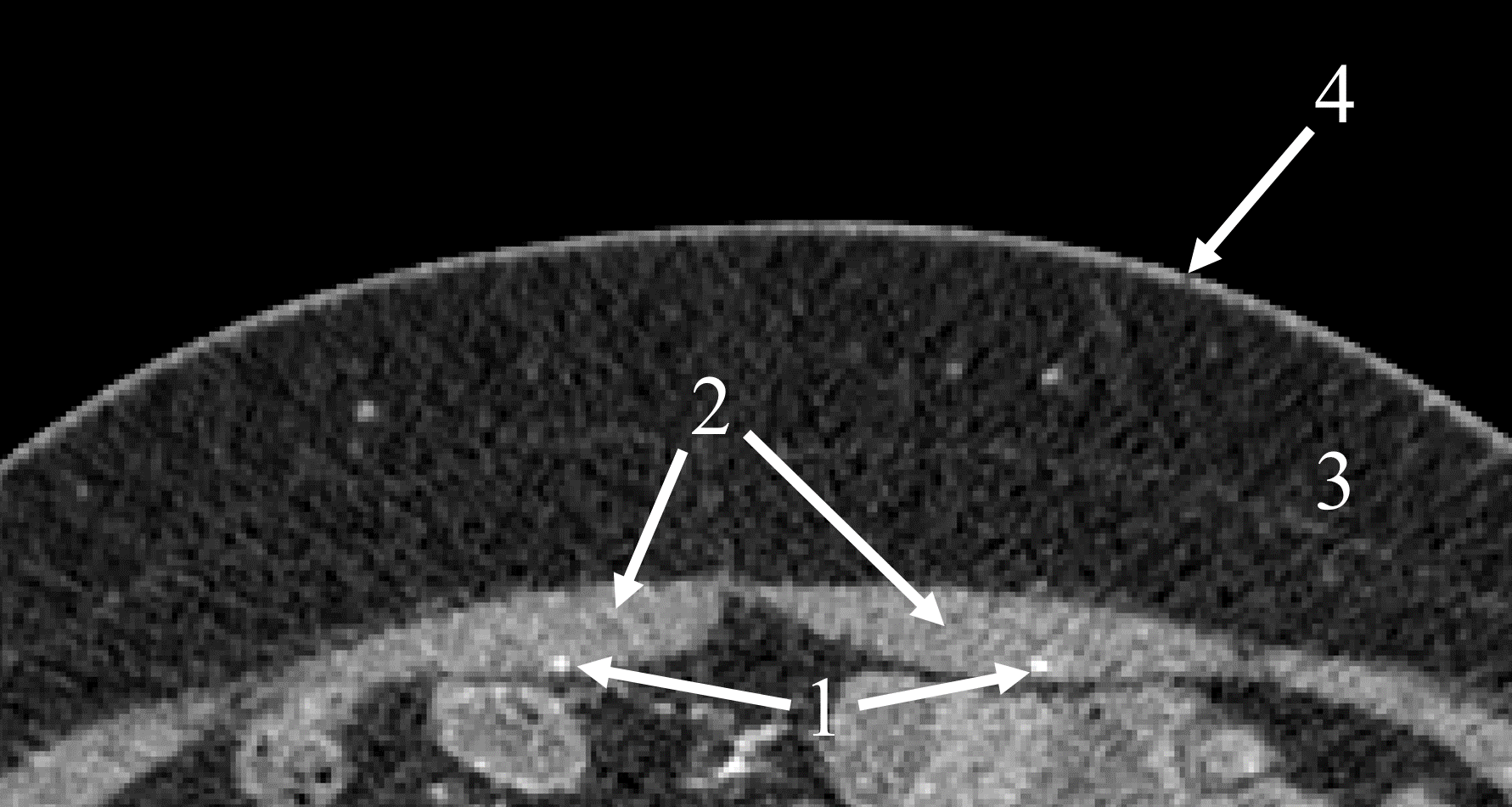}			
			\end{center}
			
		\end{minipage}
		\\ \vspace{1mm}
		\begin{minipage}{\textwidth}
			\begin{center}
				\includegraphics[height=40mm,width=40mm]{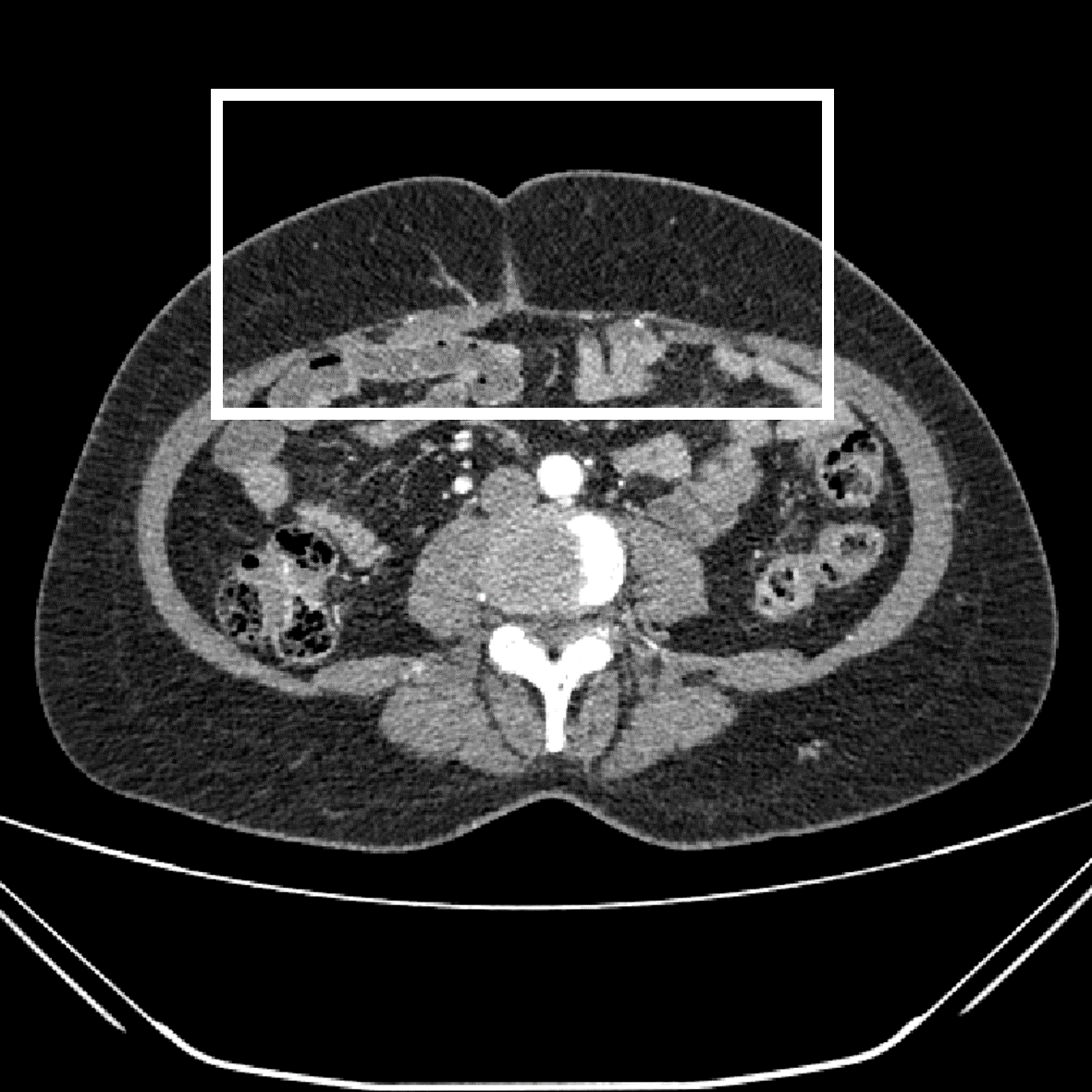}
				\includegraphics[height=40mm,width=70mm]{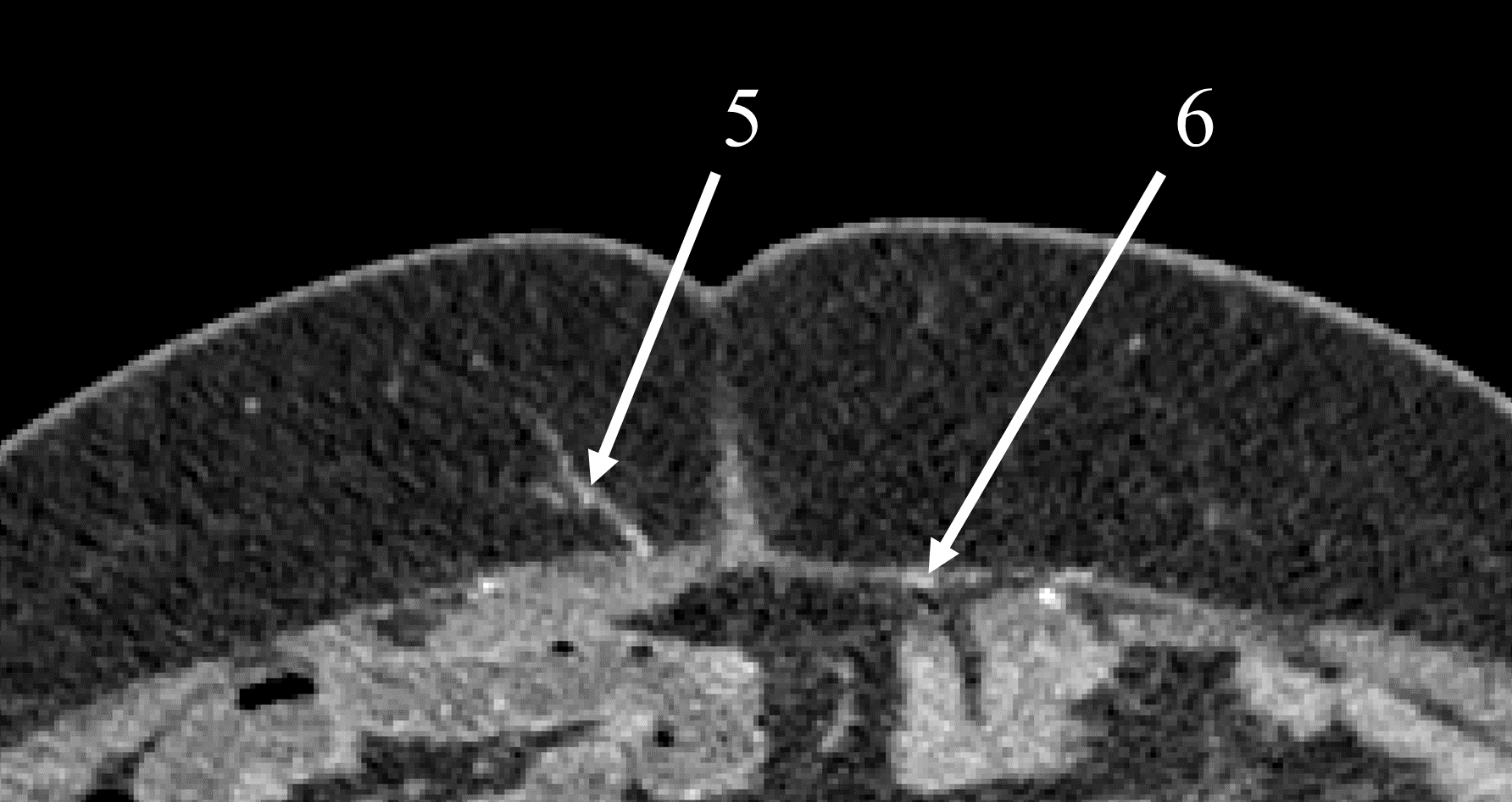}
			\end{center}
		\end{minipage}
	\end{center}
	\caption{Example axial slices of a CTA scan with the region of interest delimited by the white rectangles (left), and corresponding close up of those regions (right). The identified structures are: 1~-~DIEAs, 2~-~rectus abdominis muscle, 3~-~subcutaneous region, 4~-~skin, 5~-~subcutaneous course of a perforator, 6~-~intramuscular course of a perforator.}
	\label{fig:axi_examples}
\end{figure*}

Volumetric data of 21 patients were obtained using the CTA imaging technology, in accordance with the Declaration of Helsinki. The volumes were acquired on a Philips Brilliance Big Bore 16 slice CT scanner following a particular angiography protocol, where a contrast bolus with a volume of 120 ml and a concentration of 350 mg/ml was intravenously injected at a preferential rate of 4 ml/s. The images were taken at the ideal bolus tracking phase, as determined by the workstation. Each acquisition consists of several axial slices perpendicular to the long axis of the body, covering the entire abdominal area from the pelvic region to a little above the umbilical region. We converted the Hounsfield units (HU), a measure of radiodensity, into working units according to the parameters stored in the Digital Imaging and Communications in Medicine (DICOM) format of the volumes. Window center, window width, rescale intercept, and rescale slope had values of 60 HU, 400 HU, -1024~HU and 1 respectively. The voxel spacing differs from volume to volume. The axial voxel spacing goes from 0.55 to 0.98 mm, whereas the spacing in the long axis of the body varies from 0.40 to 1.50 mm. In the left column of Fig.~\ref{fig:axi_examples}, example axial slices acquired with the described protocol are shown.

\subsection{Volume of interest}

The CTA scans used in this study cover the entire abdominal region of the patients, however, it is known that the perforators arise in the anterior portion of the abdominal wall, such that we can consider only a small part of the entire data, which we designate as volume of interest from now on. This volume should include the location where each DIEA enters the posterior rectus sheath and the complete perforator courses. Thus, our methodology requires a manual initialization by the user, by indicating two points at the end of each perforator and the locations where the DIEAs perforate the posterior rectus sheath. We believe such effort is minimal when compared to the current manual analysis that technicians and radiologists face. The region of interest and the structures which exist there are exemplified in Fig.~\ref{fig:axi_examples}.

\subsection{Subcutaneous course extraction}

\subsubsection{Vessel direction estimation}

The local intensity gradient vector field has been analyzed in the past for addressing the design of local vessel enhancement filters~\citep{Agam:2005}. Following that contribution, here we incorporate those principles in a tracking procedure. Given a gradient vector field inside a local window, we estimate the local vessel direction $\mathbf{v}_i$ as the direction which minimizes the squared projection of the local gradient vectors into $\mathbf{v}_i$: 

\begin{equation} \label{eq:agam_eq}
E(\mathbf{v}_i)=\frac{1}{n} \sum_{k=1}^{n}(\mathbf{g}_k^T\mathbf{v}_i)^2 = \mathbf{v}_i^T \bigg( \frac{1}{n} \sum_{k=1}^{n} \mathbf{g}_k \mathbf{g}_k^T \bigg) \mathbf{v}_i
\end{equation}

\noindent
where $n$ is the number of gradient vectors inside the local window, and $g_k$ is the $kth$ gradient vector. By denoting $G\equiv(1/\sqrt n)[g_k,\ldots,g_n]$, \eqref{eq:agam_eq} becomes:

\begin{equation}\label{eq:vess_dir_simpl}
E(\mathbf{v})\equiv\mathbf{v}_i^T GG^T \mathbf{v}_i
\end{equation}

\noindent
where $GG^T$ is a $3\times3$ correlation matrix.  As proven by \citet{Agam:2005} the minimum of~\eqref{eq:vess_dir_simpl} is obtained by the eigenvector of $GG^T$ having smallest eigenvalue. In the ideal vessel case, the gradient vectors are normal to the long axis of the vessel. In the non-ideal scenario, noise will distort the gradient vector field. Even then, for adequate sizes of the local window, it is expected that the underlying vessel signal has the biggest contribution to the gradient vector field structure.

Nonetheless, since vessels spawn across very few voxels, it is unreliable to directly use intensity information to obtain the gradient vector field. Thus, we propose that the gradient field is obtained from a vessel enhanced volume, by using Frangi's vesselness~\citep{Frangi:1998} measure:

\begin{equation}\label{eq:frangi_vess}
\nu(s) = \begin{cases}
0 \hfill \text{if $\lambda_2>0$ or $\lambda_3>0$}\\
\Big(1-\me^{\frac{-{R_A}^2}{2\alpha^2}} \Big) \cdot \me^{\frac{-{R_B}^2}{2\beta^2}} \cdot \Big(1-\me^{\frac{-S^2}{2c^2}} \Big)
\end{cases}
\end{equation}

\noindent
where $s$ is a given voxel of our data domain, and $\lambda_1, \lambda_2, \lambda_3$, are eigenvalues of increasing absolute value of the Hessian matrix. The constants $\alpha$, $\beta$, and $c$ control the sensitivity of the vesselness function to the terms $R_A$, $R_B$, and $S$, which are eigenvalue-based ratios accounting for, respectively, the distinction between line-like and plate-like structures, the deviation from a blob, and the amount of structure present: 

\begin{equation}
R_A = \frac{|\lambda_2|}{|\lambda_3|}
\end{equation}

\begin{equation}
R_B = \frac{|\lambda_1|}{\sqrt{|\lambda_2\lambda_3|}}
\end{equation}

\begin{equation}
S = \sqrt{\lambda_1^2+\lambda_2^2+\lambda_3^2}
\end{equation}

\subsubsection{Ridge-based correction}

\begin{figure*}
	\begin{minipage}{1\textwidth}
		\begin{center}
			\begin{minipage}{0.24\textwidth}
				\begin{center}
					\includegraphics[width=1\textwidth]{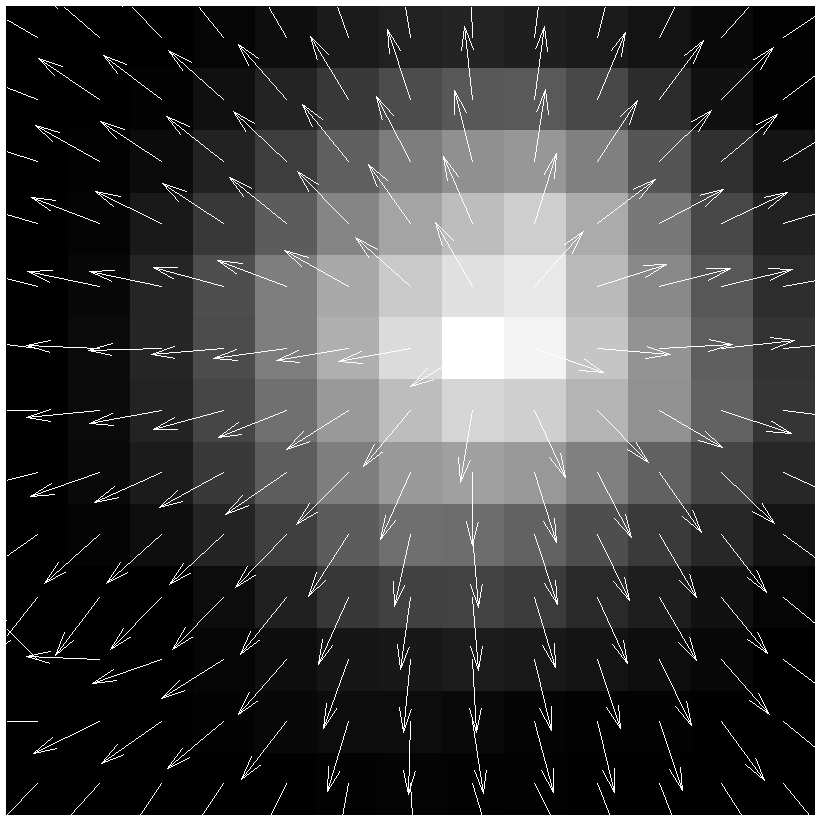}
				\end{center}
			\end{minipage}
			\begin{minipage}{0.24\textwidth}
				\begin{center}
					\includegraphics[width=1\textwidth]{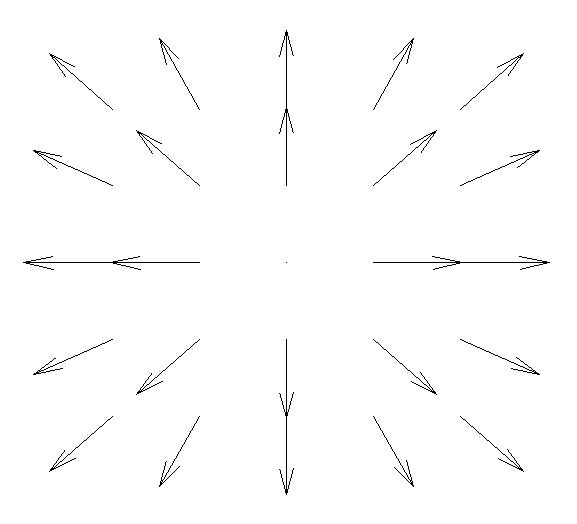}
				\end{center}	
			\end{minipage}	
			\begin{minipage}{0.24\textwidth}
				\begin{center}
					\includegraphics[width=1\textwidth]{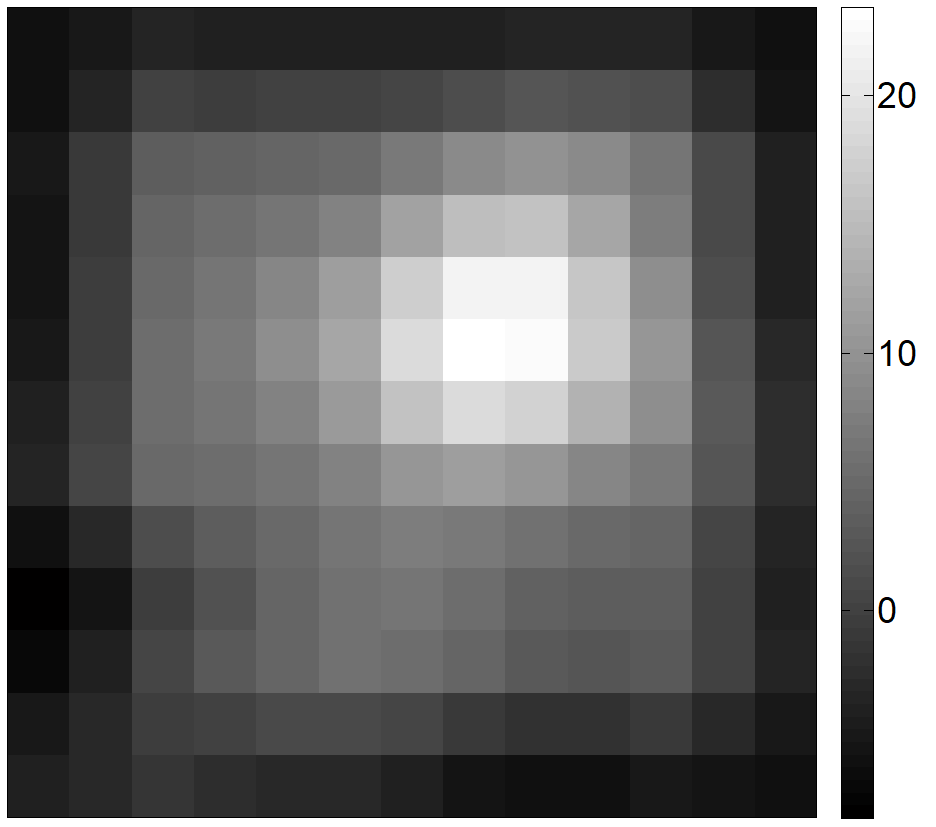}
				\end{center}
			\end{minipage}
			\begin{minipage}{0.24\textwidth}
				\begin{center}
					\includegraphics[width=1\textwidth]{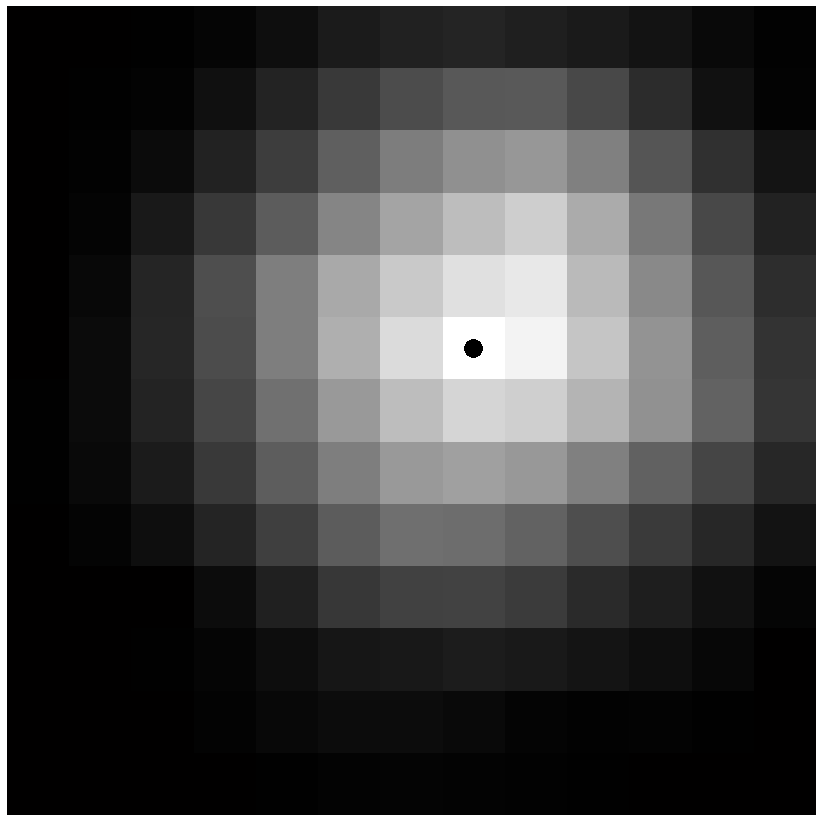}
				\end{center}	
			\end{minipage}
		\end{center}
	\end{minipage}

	\caption{Ridge-based correction framework. From left to right: example cross sectional image with gradient vector field imposed; template for finding the center; cross-correlation result; and the detected ridge. Images are interpolated for improved visualization.}
	\label{fig:correction}
\end{figure*}

Although the methodology described until this point allows to find the local direction of a vessel which goes through the window, it does not guarantee that the estimated centerline point is near the center of the vessel. To address this problem, an additional measure is taken every $N$ iterations, which is responsible for correcting deviations to the center of the vessel due to error accumulation. It relies on the assumption that voxels on the center of the vessel have higher intensity in the vessel enhanced data, and that it decreases as the distance to the center increases. In a 2D image of the cross section of a vessel, it is then expected that the center location can be found by analyzing the divergence of the gradient vector field. After predicting the position of the new centerline point $\tilde{X}_{i+1}$ using the local gradient vectors information, the plane which contains that point and is orthogonal to the vessel direction $\mathbf{v}_i$ is obtained. It is expected that this plane includes a roughly circular brighter region which is the 2D cross section of the vessel enhanced data (see the left image in Figure~\ref{fig:correction}). The gradient vector field is calculated~\citep{Oliveira:2014} and its similarity to the template represented in Fig.~\ref{fig:correction} is assessed through cross-correlation:

\begin{equation}
(\mathbf{f}*\mathbf{g})[\eta] \myeq \sum_m \mathbf{f}^*[m]\mathbf{g}[\eta+m]
\end{equation}

where $\mathbf{f}$ and $\mathbf{g}$ represent the gradient orientation vector field and template vector field, respectively, $\mathbf{f}^*$ is the complex conjugate of $\mathbf{f}$, and $\eta$ is the displacement. The center location estimation corresponds to the maximum response location (see the right image in Fig.~\ref{fig:correction}).

\subsubsection{Smoothness properties}

\begin{figure}
	\begin{center}
		\includegraphics[width=6cm]{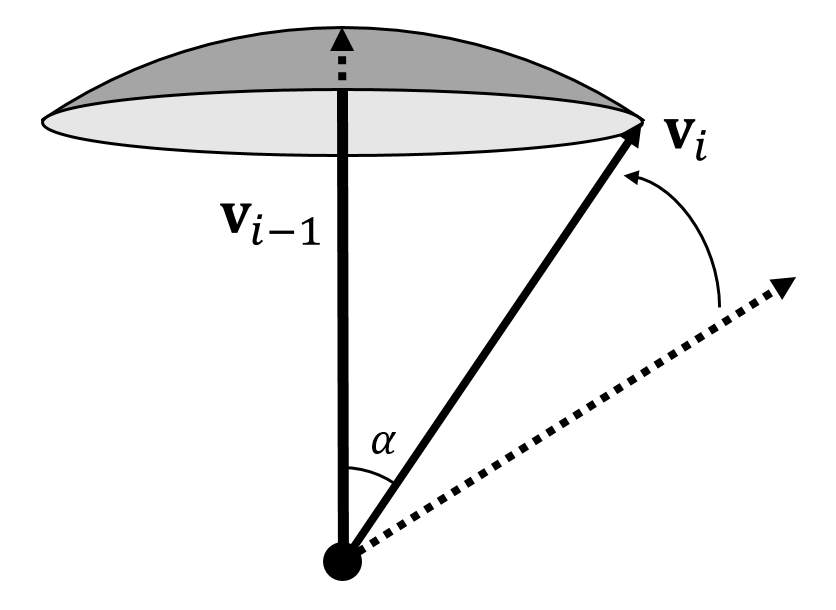}
	\end{center}
	\caption{Direction variation restriction between consecutive iterations. $\alpha$ controls the smoothing factor.}
	\label{fig:smooth}
\end{figure}

To assure smoothness throughout the tracking procedure, one option would be to use a Kalman filter~\citep{Kalman:1960}. This approach would allow to fuse the estimates of the direction estimator and the ridge-based corrector, and also tune the degree of belief in each of them. Even then, we found it was sufficient to simply set an upper bound on the direction variation allowed at consecutive iterations. This may be interpreted as obligating $\mathbf{v}_i$ to lie on a spherical cap centered at $\mathbf{v}_{i-1}$, as represented in Fig.~\ref{fig:smooth}. The higher we allow the direction variation to be, the larger is the spherical cap.

\subsection{Intramuscular course extraction}

The intramuscular course of the perforators is commonly hard to detect as the muscle signal is also significant in a CTA acquisition, turning general tracking procedures not adequate for this task. Hence, we propose the use of a minimum cost path method to find the intramuscular vessel pathway between the site where the perforator reaches the fascia and the manually given DIEA landmark. Thus, the problem becomes constrained to finding a path that connects two voxels, leading to a decrease in the computational effort required. In this work, we propose the use of the A* search algorithm~\citep{Hart:1968}, as it includes a heuristic to improve the search performance. At each iteration, the A* search algorithm expands the path which minimizes the following expression:

\begin{equation}
f(s)=g(s)+h(s)
\end{equation}

\noindent
where $s$ is the last node on the path, $g(s)$ is the cost of the path from the start node to $s$, and $h(s)$ is the heuristic that estimates the cost of the cheapest path from $s$ to the goal. In this work, the Euclidean distance between $s$ and the target voxel is used as the heuristic function. The cost of travelling from one node to another is given by the following expression:

\begin{equation}
c_{s_1,s_2}=C(s_2) \cdot d(s_1,s_2)
\end{equation}

\noindent
where $s_1$ is the current node, $s_2$ is a neighbor node, $d(s_1,s_2)$ is the Euclidean distance between those nodes, and $C(s_2)$ is the terrain cost of the neighbor node.

To find the desired pathways, lower costs must be given to vessel voxels. In this work, we analyze the usefulness of a function that has been used for the detection of coronary arteries \citep{Metz:2009}:

\begin{equation} \label{eq:family2}
C(s) = \frac{1}{\nu_N(s) \cdot T(s)+\epsilon}
\end{equation}

\noindent
where $\nu_N(s)$ is Frangi's Vesselness~\eqref{eq:frangi_vess} at voxel $s$ normalized to the range $\left[0,1\right]$, $\epsilon$ is a small constant to avoid division by zero, and $T(s)$ is a sigmoid function of the intensity:

\begin{equation} \label{eq:sig_int}
T(s) = \frac{1}{1+\me^{a_s(I(s)-b_s)}}
\end{equation}

\noindent
with $I(s)$ being the intensity at voxel $s$, and $a_s$ and $b_s$ constants controlling the shape of the sigmoid function.

The expression~\eqref{eq:family2} produces low costs at voxels which have high probability of belonging to a vessel, according to~\eqref{eq:frangi_vess}, and which have relatively high radiodensity, according to the parameterization of~\eqref{eq:sig_int}. Note that~\eqref{eq:family2} gives costs in the range $\left[1,\infty\right[$, guaranteeing that the heuristic of the A* is admissible.

\section{Results and Discussion} \label{sec:results}

In order to evaluate the proposed methodology, a radiologist provided manual annotations of the existing perforators in the database described in section~\ref{sec:methods}, by defining some landmarks belonging to the centerlines of those vessels.  Across the 21 volumes, a total of 98 subcutaneous and 50 intramuscular perforator pathways were identified. Since the annotations are sparse when compared to the extracted paths, we consider the Euclidean and Hausdorff distances from the \textit{Ground Truth} annotations to the extracted paths, as metrics indicating the precision of the vessel detection methodologies. The Euclidean distance measures the average distance from the path to the manual annotation, whereas the Hausdorff distance accounts for the maximum distance. Additionally, as the minimum path approach designed for intramuscular path extraction may take a noticeable amount of time to run, we also evaluate the time efficiency of the proposed methodology for that task.

\subsection{Subcutaneous course extraction}

Our proposed methodology for extracting the subcutaneous portion of perforators is now assessed in terms of the metrics described before, and compared to the tracking algorithm of \citet{Friman:2010}, which was tailored for the robust detection of small vessels.

Regarding the implementation of our approach, we empirically set the step $\delta$ to 1 mm, the side length of the local window to 4 mm, took a correction measure every 3 iterations, restricted the direction variation to 60 degrees, and obtained the vessel enhanced data~\eqref{eq:frangi_vess} with $\alpha=0.5$, $\beta=10$, and $c=500$. In order to obtain the results from the approach of \citet{Friman:2010}, we used their implementation in MeVisLab~\citep{Mevislab}, and tuned the parameters for this particular application, by setting the minimum and maximum radius to 0.5 and 1.5 mm, respectively, the step length to 1 mm, and the maximum step angle to 60 degrees. The initialization of both perforator tracking procedures was made at the \textit{Ground Truth} landmark which was closer to the end of the perforator. Table~\ref{results_sub} summarizes our findings.

\begin{table}[t!]
	\caption{Results of the proposed method and the approach of~\citet{Friman:2010}, concerning the subcutaneous course extraction, when comparing to the ground truth annotations.}
	\begin{center}
		\begin{tabular}{|c|c|c|}
			\hline
			\multirow{2}{*}{\textbf{Method}}&\multicolumn{2}{|c|}{\textbf{Path error} (mm)} \\
			\cline{2-3} 
			& \textbf{Euclidean distance}& \textbf{Hausdorff distance}  \\
			\hline
			proposed & 0.64 $\pm$ 0.25 & 1.17 $\pm$ 0.88 \\
			\hline
			\citet{Friman:2010} & 1.01 $\pm$ 0.60  & 2.38 $\pm$ 2.17 \\
			\hline
		\end{tabular}
		\label{results_sub}
	\end{center}
\end{table}

\begin{figure*}
	\begin{center}
	\begin{minipage}{0.48\textwidth}
		\begin{center}
			\includegraphics[width=5.6cm]{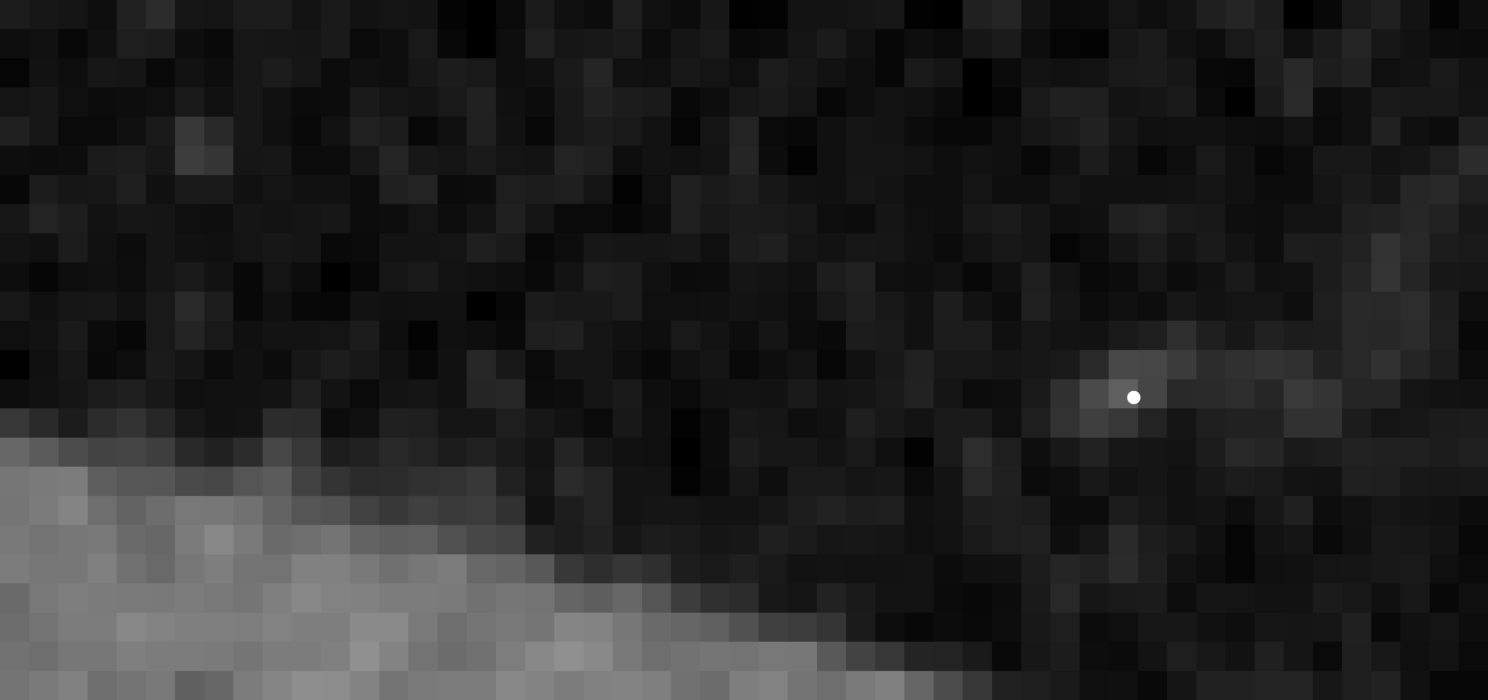}
			
			\vspace{1mm}
			
			\includegraphics[width=5.6cm]{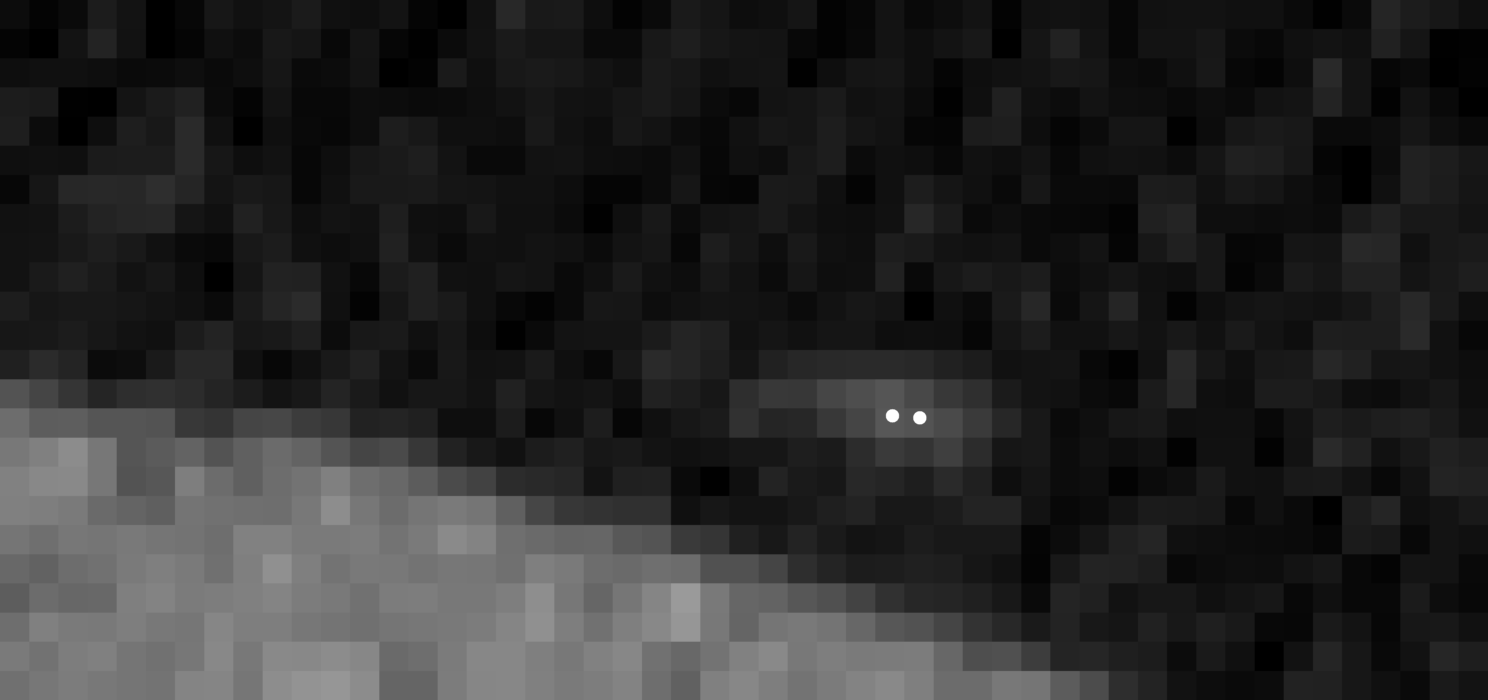}
			
			\vspace{1mm}
			
			\includegraphics[width=5.6cm]{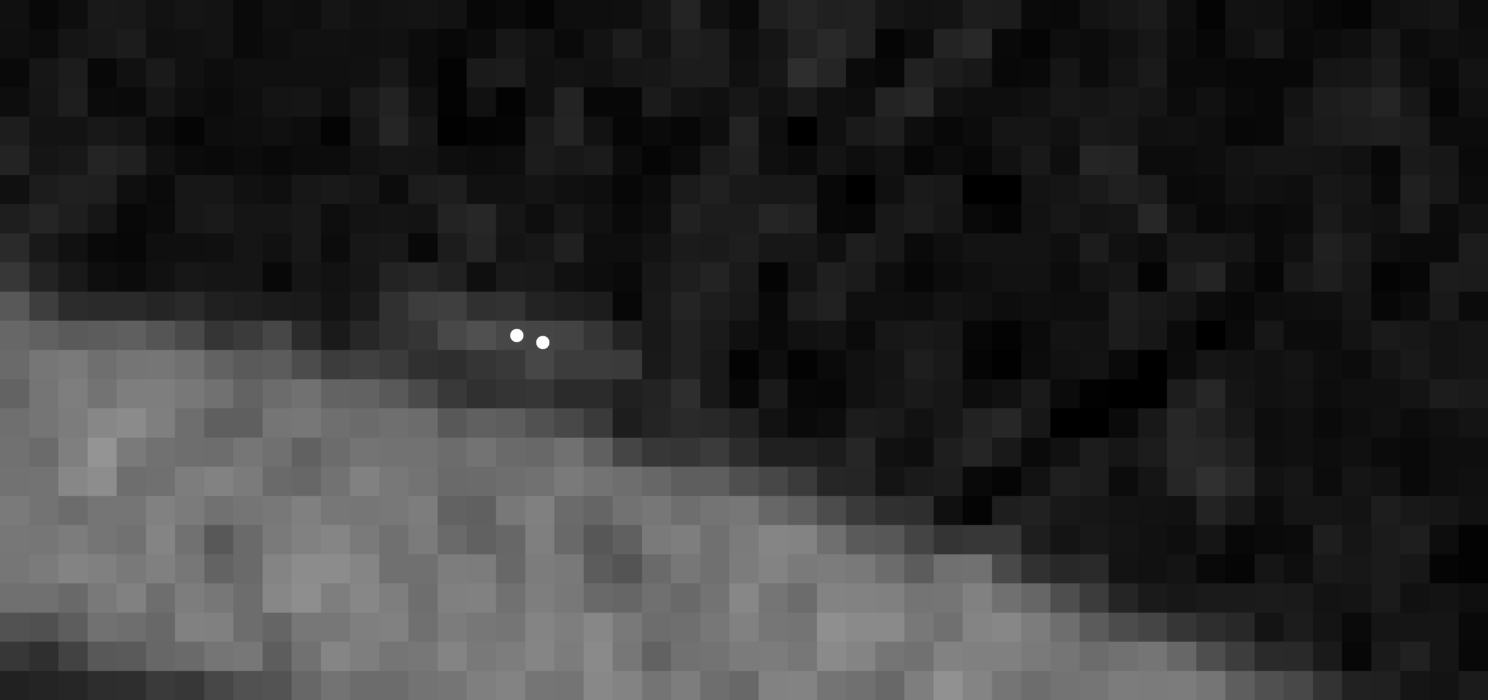}
			
			\vspace{1mm}
			
			\includegraphics[width=5.6cm]{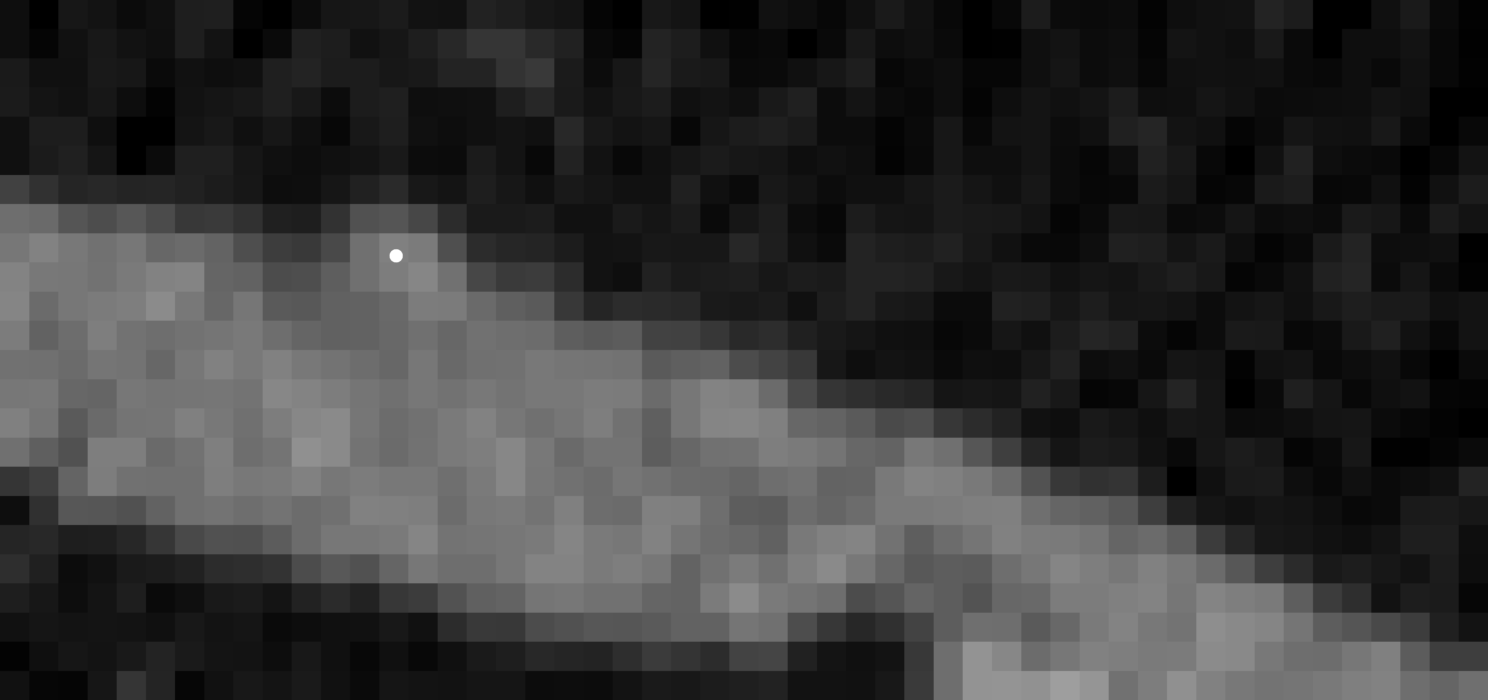}
		\end{center}		
	\end{minipage}
	\begin{minipage}{0.48\textwidth}
		\begin{center}
			\includegraphics[width=5.6cm]{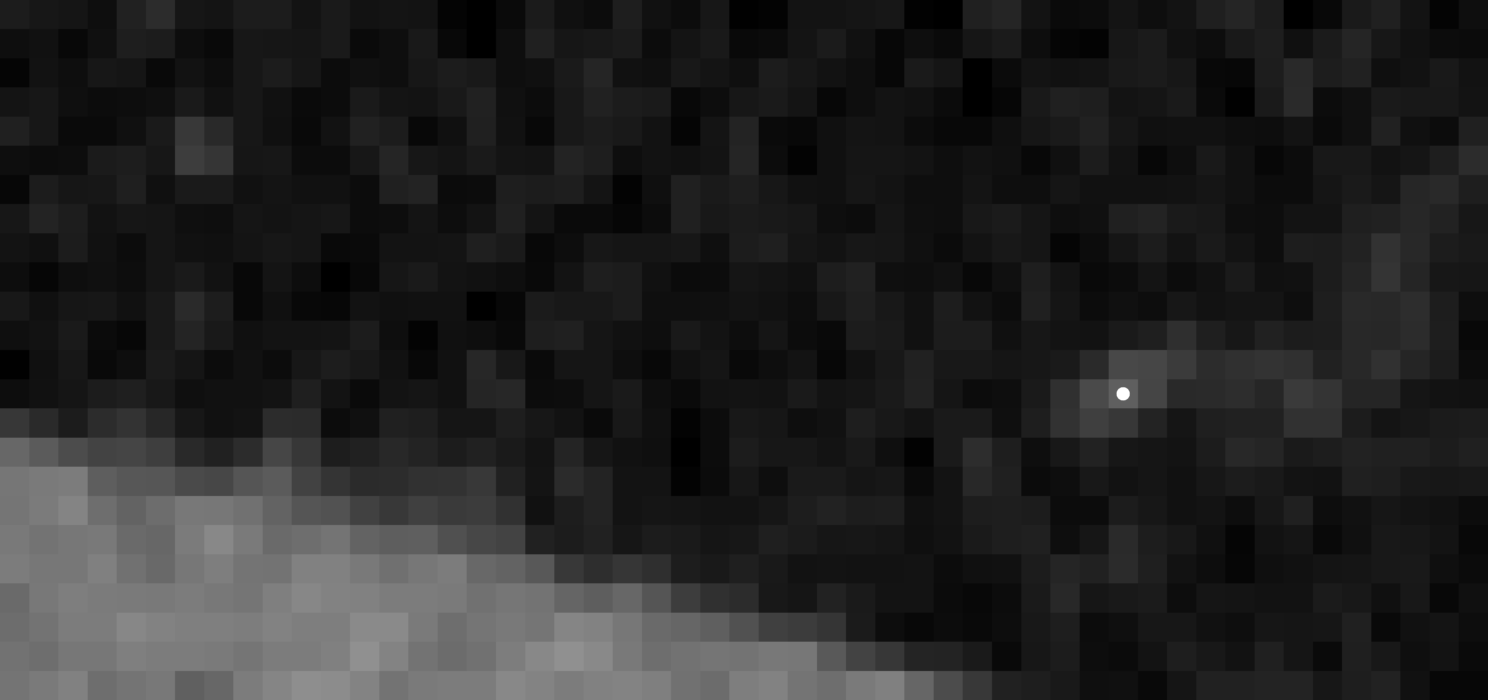}
			
			\vspace{1mm}
			
			\includegraphics[width=5.6cm]{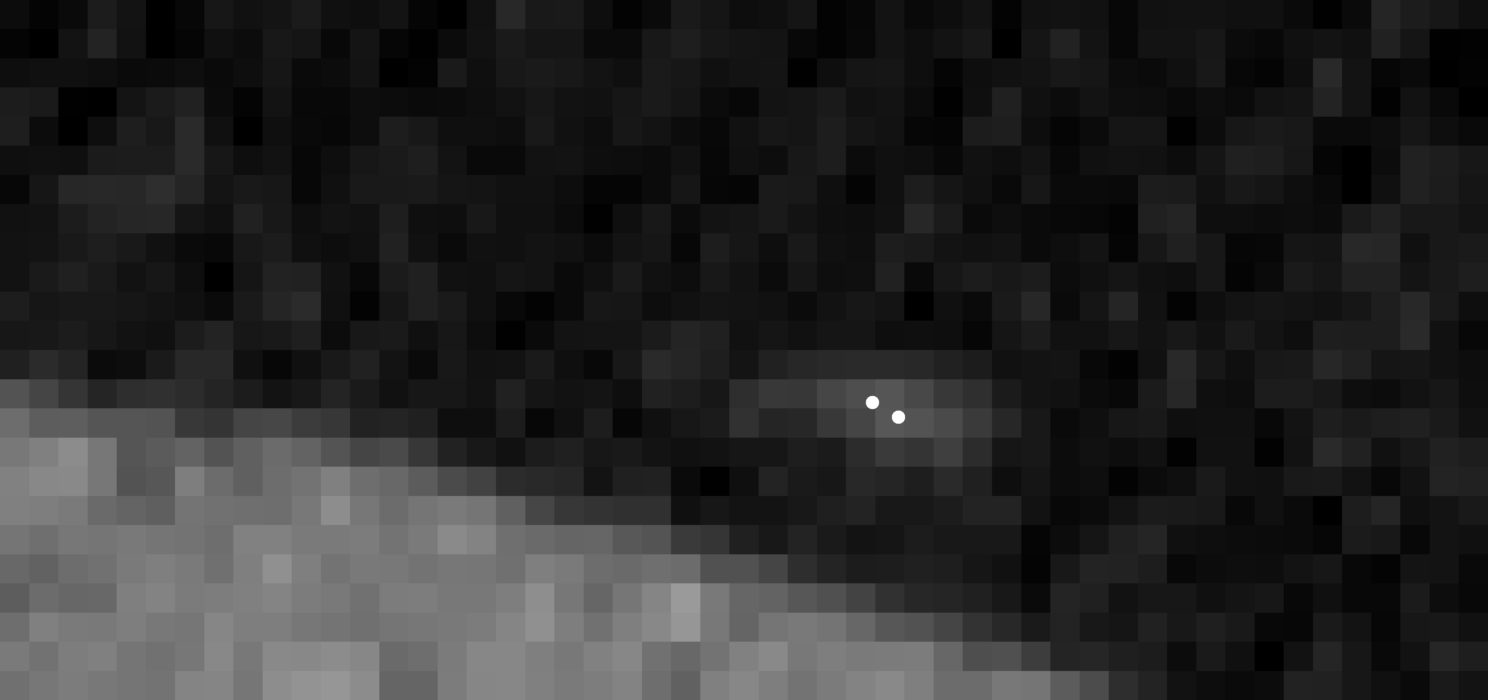}
			
			\vspace{1mm}
			
			\includegraphics[width=5.6cm]{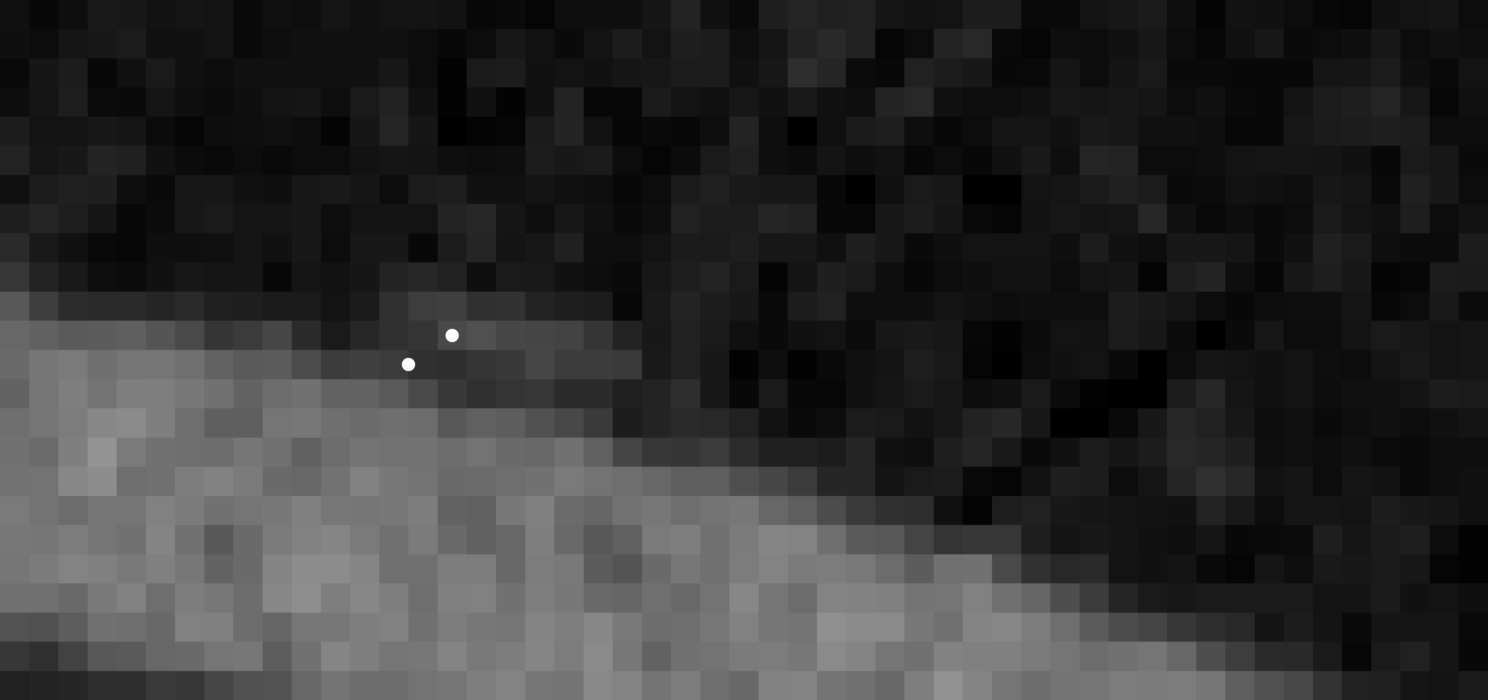}
			
			\vspace{1mm}
			
			\includegraphics[width=5.6cm]{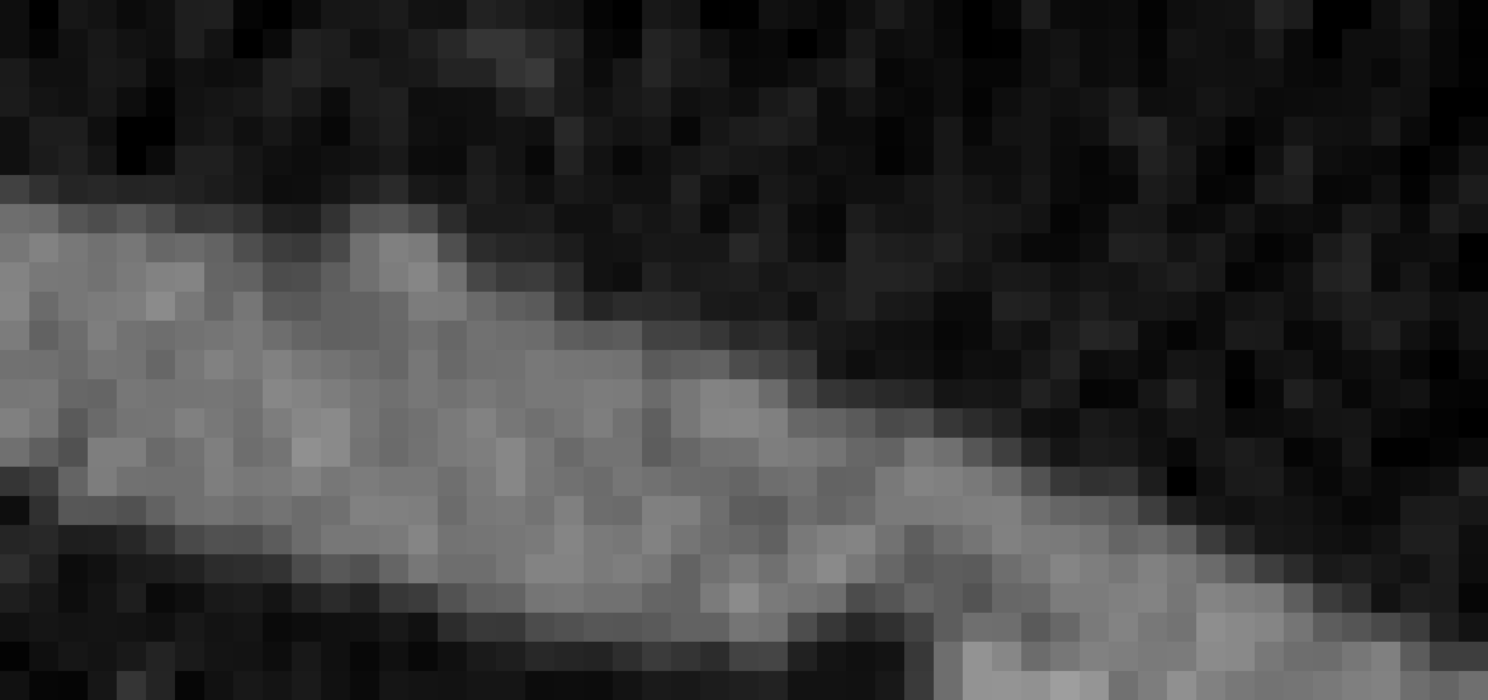}
		\end{center}
	\end{minipage}
	\caption{Tracking the subcutaneous course of a perforator using our approach (left) and the method proposed by \citet{Friman:2010} (right). The latter is more prone to terminating sooner when the vessel evolves near the fascia.}
	\label{fig:subcut_detections}
	\end{center}

\end{figure*}

Our methodology proved to be more adequate to the subcutaneous tracking of perforators, as the average Euclidean and Hausdorff distances were significantly lower when using it. In fact, our methodology reached subvoxel accuracy for most of the volumes in our database. The increased performance was mainly due to the fact that our approach is able to neglect more the presence of the muscle when tracking the vessel, as we do it in the vessel enhanced data. In contrast, the methodology of Friman et al. suffered more from this. The average Hausdorff distance gives information about how well the methods are able to correctly track the perforator until it reaches the fascia, as it is near this region that the tracking procedure faces more difficulties, especially when the perforator has a substancial overlap with the muscle signal. Again, the proposed method behaved better in such circumstance, thus being more adequate for determining the location where the perforators leave the fascia. Fig.~\ref{fig:subcut_detections} shows a comparison between the methods in an example subcutaneous path having a segment close to the fascia.

\subsection{Intramuscular course extraction}

\begin{figure*}[t]
		\begin{center}
		\begin{minipage}{0.49\textwidth}
			\begin{center}
				\includegraphics[height=5cm]{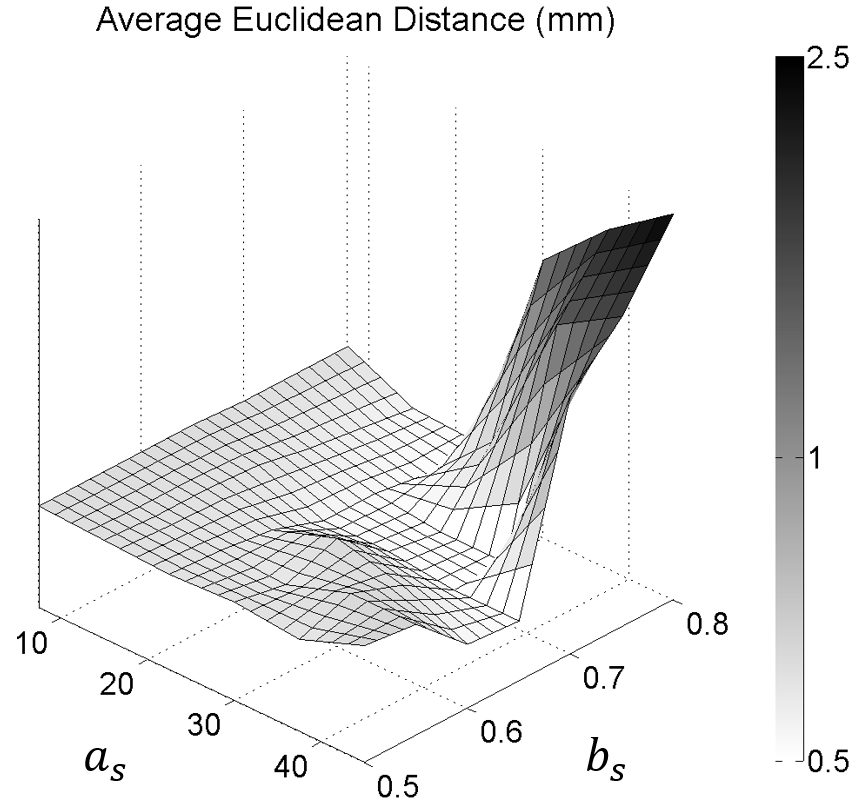}	
			\end{center}		
		\end{minipage}	
		\begin{minipage}{0.49\textwidth}
			\begin{center}
				\includegraphics[height=5cm]{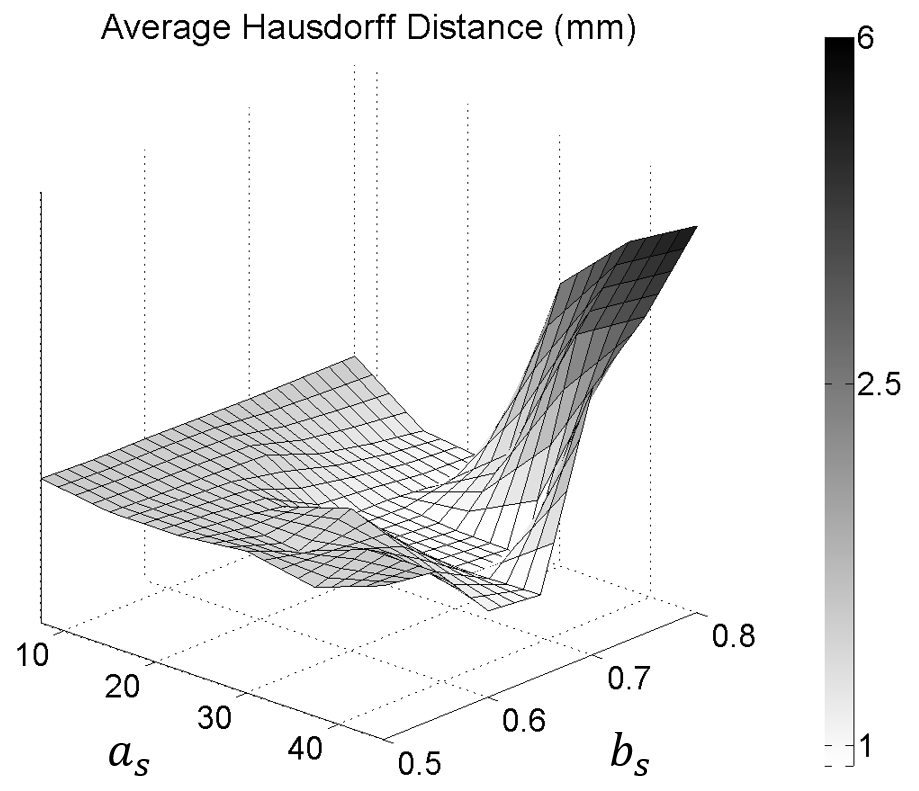}		
			\end{center}		
		\end{minipage}

	    \vspace{1cm}

		\begin{minipage}{0.49\textwidth}
			\begin{center}
				\includegraphics[height=5cm]{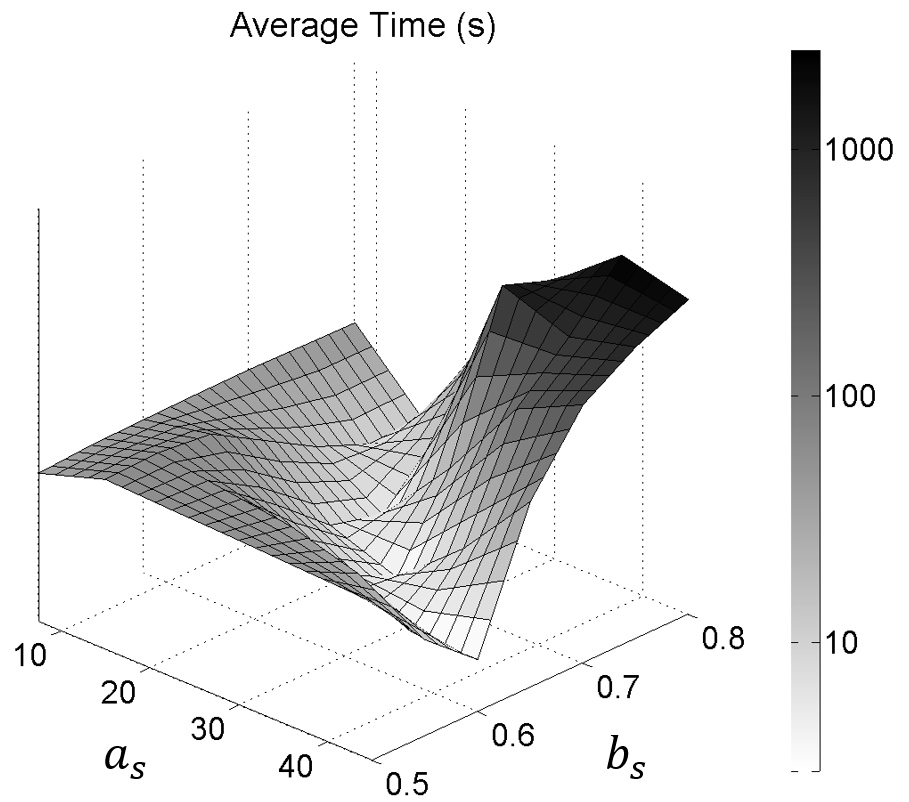}		
			\end{center}	
		\end{minipage}
		\caption{Manifolds showing the average A* performance when using~\eqref{eq:family2}, and how it varies according to the parameterization ($a_s$, $b_s$). Logarithmic scales were used for the sake of clarity.}
		\label{fig:mp_manifolds}
		\end{center}
			
	\end{figure*}

Concerning the extraction of intramuscular paths, we seek to find how appropriate is the volume given by~\eqref{eq:family2} when used as terrain costs inside a minimum path framework such as A*. Note that by appropriate, we refer to path accuracy but also time expended, as minimum path approaches may become prohibitively slow when there is a need to visit many neighbors, especially in a 3D environment. To obtain $\nu_N(s)$, we empirically set $\alpha=0.5$, $\beta =0.5 $, and $c = 100$. Regarding $T(s)$, we considered 42 different parameterizations, as given by the possible combinations of taking $a_s$ from $7.5$ to $45$ with a step of $7.5$, and $b_s$ from $0.5$ to $0.8$ with a step of $0.05$ (radiodensity was mapped into the range $\left[0,1\right]$ as described in section~\ref{sec:methods}). Besides the metrics considered in the case of subcutaneous extraction evaluation, Euclidean and Hausdorff distance between the \textit{Ground Truth} annotations and the retrieved paths, we also measured the time expended. Table~\ref{results_intra_2} presents the results of our experiments. For the sake of readability, we show only the average value of the metrics of interest. The performance manifold, concerning each of the metrics, is visually represented in Figure~\ref{fig:mp_manifolds}.

\begin{table}[]
	\centering
	\caption{A* performance when retrieving the intramuscular courses of the perforators, using~\eqref{eq:family2} to obtain the volume of costs. Each cell contains the performance of a parameterization ($a_s$, $b_s$) regarding the average Euclidean and Hausdorff distances between the \textit{Ground Truth} annotations and the retrieved paths (mm), and the average time expended (s), respectively.}
	\label{results_intra_2}
	\begin{tabular}{@{}ccccccccc@{}}
		\multicolumn{1}{l}{} & \multicolumn{1}{l}{}      & \multicolumn{7}{c}{$b_s$} \\
		\multicolumn{1}{l}{} & \multicolumn{1}{l}{}      & 0.50                                                                            & 0.55                                                                            & 0.60                                                                                                                           & 0.65                                                                                                                           & 0.70                                                                            & 0.75                                                                            & 0.80                                                                            \\ \cmidrule(l){3-9} 
		\multirow{6}{*}{$a_s$} & \multicolumn{1}{c|}{7.5}  & \multicolumn{1}{c|}{\begin{tabular}[c]{@{}c@{}}0.62\\ 1.40\\ 36.1\end{tabular}} & \multicolumn{1}{c|}{\begin{tabular}[c]{@{}c@{}}0.62\\ 1.38\\ 39.5\end{tabular}} & \multicolumn{1}{c|}{\begin{tabular}[c]{@{}c@{}}0.62\\ 1.38\\ 40.8\end{tabular}}                                                & \multicolumn{1}{c|}{\begin{tabular}[c]{@{}c@{}}0.62\\ 1.37\\ 41.8\end{tabular}}                                                & \multicolumn{1}{c|}{\begin{tabular}[c]{@{}c@{}}0.61\\ 1.36\\ 42.2\end{tabular}} & \multicolumn{1}{c|}{\begin{tabular}[c]{@{}c@{}}0.61\\ 1.36\\ 41.1\end{tabular}} & \multicolumn{1}{c|}{\begin{tabular}[c]{@{}c@{}}0.61\\ 1.36\\ 40.1\end{tabular}} \\ \cmidrule(l){3-9} 
		& \multicolumn{1}{c|}{15}   & \multicolumn{1}{c|}{\begin{tabular}[c]{@{}c@{}}0.61\\ 1.32\\ 60.8\end{tabular}} & \multicolumn{1}{c|}{\begin{tabular}[c]{@{}c@{}}0.60\\ 1.30\\ 64.2\end{tabular}} & \multicolumn{1}{c|}{\begin{tabular}[c]{@{}c@{}}0.60\\ 1.29\\ 56.1\end{tabular}}                                                & \multicolumn{1}{c|}{\begin{tabular}[c]{@{}c@{}}0.59\\ 1.27\\ 26.7\end{tabular}}                                                & \multicolumn{1}{c|}{\begin{tabular}[c]{@{}c@{}}0.56\\ 1.13\\ 12.7\end{tabular}} & \multicolumn{1}{c|}{\begin{tabular}[c]{@{}c@{}}0.54\\ 1.06\\ 7.48\end{tabular}} & \multicolumn{1}{c|}{\begin{tabular}[c]{@{}c@{}}0.52\\ 1.01\\ 7.90\end{tabular}} \\ \cmidrule(l){3-9} 
		& \multicolumn{1}{c|}{22.5} & \multicolumn{1}{c|}{\begin{tabular}[c]{@{}c@{}}0.60\\ 1.32\\ 56.7\end{tabular}} & \multicolumn{1}{c|}{\begin{tabular}[c]{@{}c@{}}0.59\\ 1.28\\ 51.2\end{tabular}} & \multicolumn{1}{c|}{\begin{tabular}[c]{@{}c@{}}0.57\\ 1.22\\ 19.6\end{tabular}}                                                & \multicolumn{1}{c|}{\begin{tabular}[c]{@{}c@{}}0.52\\ 0.99\\ 6.70\end{tabular}}                                                & \multicolumn{1}{c|}{\begin{tabular}[c]{@{}c@{}}0.52\\ 0.99\\ 4.80\end{tabular}} & \multicolumn{1}{c|}{\begin{tabular}[c]{@{}c@{}}0.50\\ 0.96\\ 17.1\end{tabular}} & \multicolumn{1}{c|}{\begin{tabular}[c]{@{}c@{}}0.51\\ 0.96\\ 109\end{tabular}}  \\ \cmidrule(l){3-9} 
		& \multicolumn{1}{c|}{30}   & \multicolumn{1}{c|}{\begin{tabular}[c]{@{}c@{}}0.61\\ 1.38\\ 53.4\end{tabular}} & \multicolumn{1}{c|}{\begin{tabular}[c]{@{}c@{}}0.58\\ 1.22\\ 41.3\end{tabular}} & \multicolumn{1}{c|}{\begin{tabular}[c]{@{}c@{}}0.64\\ 1.38\\ 7.10\end{tabular}}                                                & \multicolumn{1}{c|}{\begin{tabular}[c]{@{}c@{}}0.52\\ 0.98\\ 3.70\end{tabular}}                                                & \multicolumn{1}{c|}{\begin{tabular}[c]{@{}c@{}}0.50\\ 0.95\\ 28.7\end{tabular}} & \multicolumn{1}{c|}{\begin{tabular}[c]{@{}c@{}}0.69\\ 1.31\\ 1340\end{tabular}} & \multicolumn{1}{c|}{\begin{tabular}[c]{@{}c@{}}1.53\\ 3.11\\ 843\end{tabular}}  \\ \cmidrule(l){3-9} 
		& \multicolumn{1}{c|}{37.5} & \multicolumn{1}{c|}{\begin{tabular}[c]{@{}c@{}}0.59\\ 1.28\\ 49.5\end{tabular}} & \multicolumn{1}{c|}{\begin{tabular}[c]{@{}c@{}}0.65\\ 1.41\\ 25.9\end{tabular}} & \multicolumn{1}{c|}{\begin{tabular}[c]{@{}c@{}}0.59\\ 1.18\\ 3.51\end{tabular}}                                                & \multicolumn{1}{c|}{\begin{tabular}[c]{@{}c@{}}\textbf{0.50}\\ \textbf{0.96}\\ \textbf{14.6}\end{tabular}} & \multicolumn{1}{c|}{\begin{tabular}[c]{@{}c@{}}0.50\\ 0.94\\ 237\end{tabular}}  & \multicolumn{1}{c|}{\begin{tabular}[c]{@{}c@{}}1.48\\ 3.00\\ 756\end{tabular}}  & \multicolumn{1}{c|}{\begin{tabular}[c]{@{}c@{}}2.10\\ 4.54\\ 3031\end{tabular}} \\ \cmidrule(l){3-9} 
		& \multicolumn{1}{c|}{45}   & \multicolumn{1}{c|}{\begin{tabular}[c]{@{}c@{}}0.67\\ 1.48\\ 46.3\end{tabular}} & \multicolumn{1}{c|}{\begin{tabular}[c]{@{}c@{}}0.65\\ 1.42\\ 10.2\end{tabular}} & \multicolumn{1}{c|}{\begin{tabular}[c]{@{}c@{}}\textbf{0.51}\\ \textbf{1.00}\\ \textbf{3.50}\end{tabular}} & \multicolumn{1}{c|}{\begin{tabular}[c]{@{}c@{}}0.50\\ 0.97\\ 89.1\end{tabular}}                                                & \multicolumn{1}{c|}{\begin{tabular}[c]{@{}c@{}}1.39\\ 2.86\\ 514\end{tabular}}  & \multicolumn{1}{c|}{\begin{tabular}[c]{@{}c@{}}1.81\\ 3.81\\ 1154\end{tabular}} & \multicolumn{1}{c|}{\begin{tabular}[c]{@{}c@{}}2.67\\ 5.70\\ 2022\end{tabular}} \\ \cmidrule(l){3-9} 
	\end{tabular}
\end{table}

\begin{figure}
	\begin{center}
		\includegraphics[width=5cm]{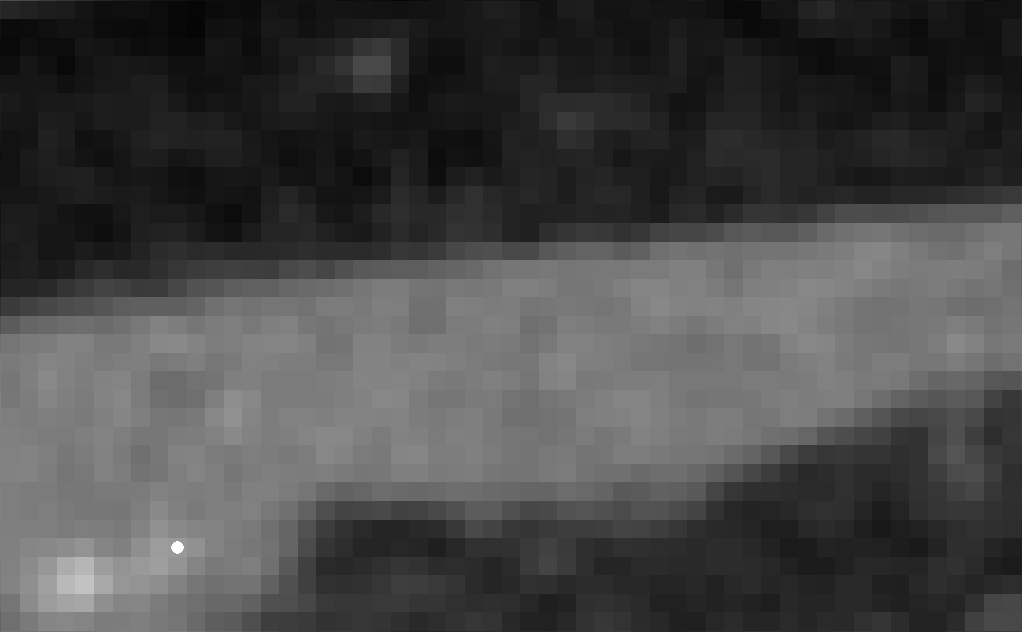}
		
		\vspace{1mm}
		
		\includegraphics[width=5cm]{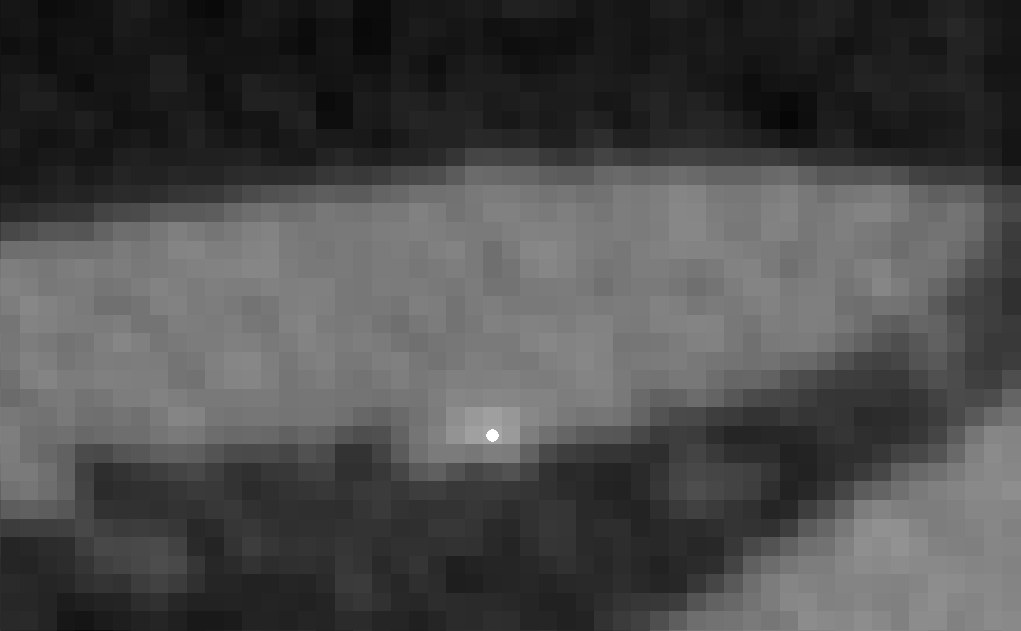}
		
		\vspace{1mm}
		
		\includegraphics[width=5cm]{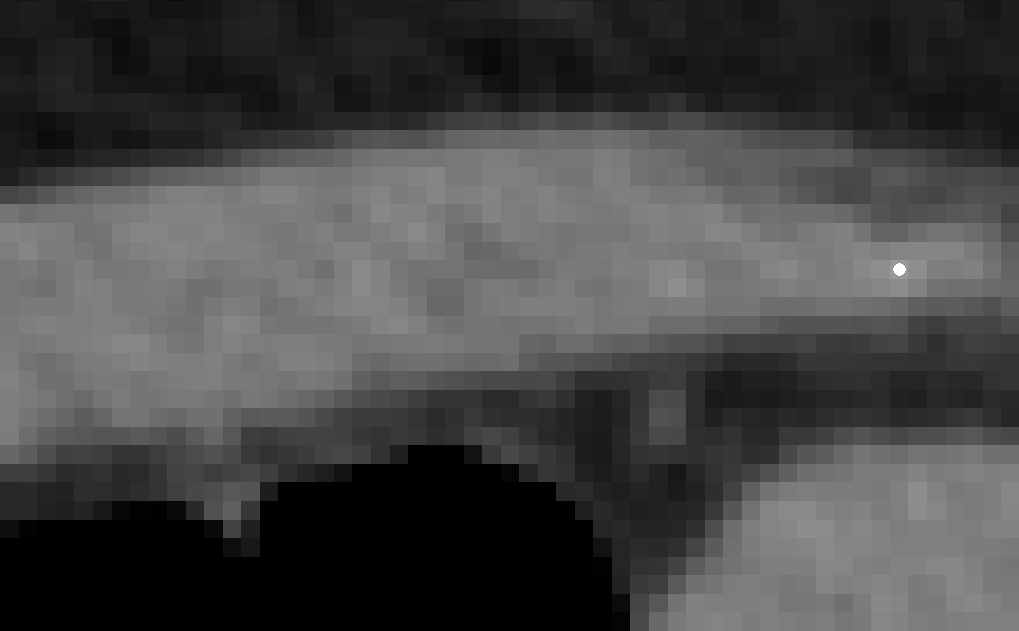}
		
		\vspace{1mm}
		
		\includegraphics[width=5cm]{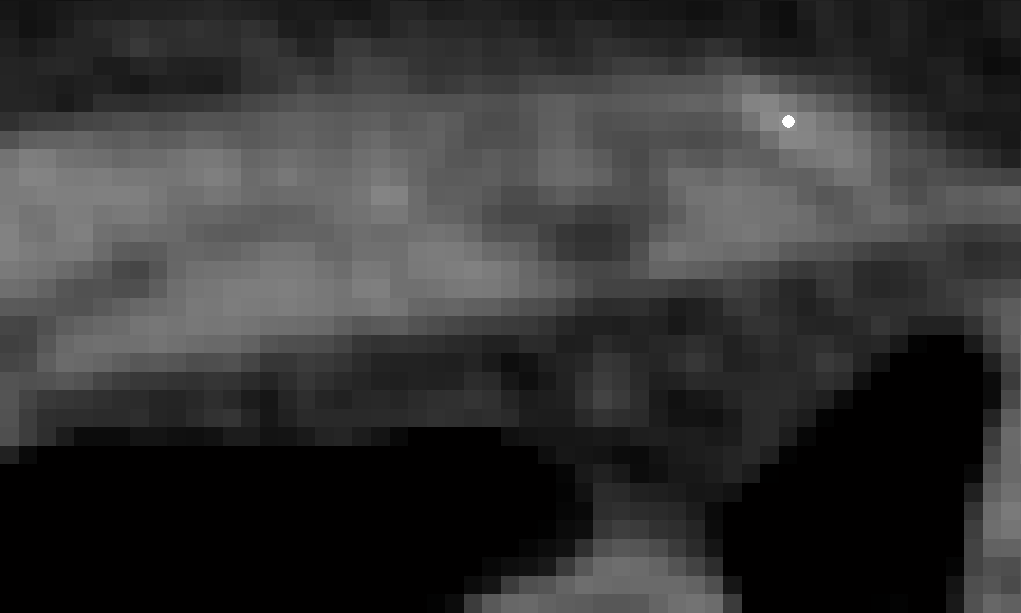}
	\end{center}
	\caption{Example intramuscular course extracted by the proposed minimum cost path method, for $a_s = 45$ and $b_s=0.60$.}
	\label{fig:intra_detections}
\end{figure}

The manifolds allow us to conclude that the influence of parameters $a_s$ and $b_s$ on the overall performance is not linear. Instead, it is a particular combination of both that may lead to a reasonable volume of costs. This was somewhat expected, as $a_s$ dictates the steepness of the sigmoid function, hence the intensity compression, and $b_s$ sets the threshold of the sigmoid, controlling the range of intensities that produce lower costs.
The parameterizations highlighted in bold in Table~\ref{results_intra_2} were the ones reaching better compromises in what concerns path detection accuracy and time expended. In a clinical setting like the one described in this paper, where the manual analysis of the data may easily reach a couple of hours, having a semi-automatic algorithm that takes a dozen of seconds to detect an intramuscular path is not problematic. Even then, our methodology was able to reach very interesting compromises. For example, the parameterization ($a_s=45$, $b_s=0.60$) was able to attain one of the best path accuracies (Euclidean and Hausdorff distances of $0.51 \pm 0.14$ mm and $1.00 \pm 0.39$ mm) and also be very fast ($3.50 \pm 6.5$ s). An example of an extracted intramuscular path using this configuration is present in Fig.~\ref{fig:intra_detections}.

\begin{figure}[t!]
		\begin{center}
			\includegraphics[width=0.8\columnwidth]{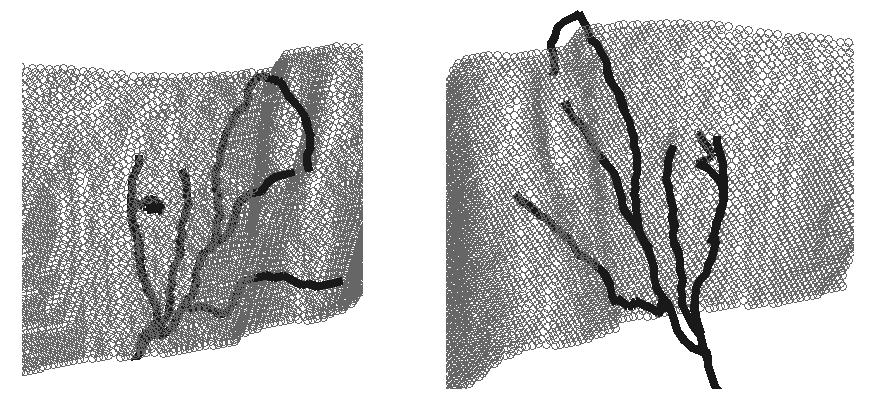}
		\end{center}
		\caption{3D representation of the fascia and the extracted vascular network of one hemiabdomen.}
		\label{fig:render_3d}
	\end{figure}

Thus, for appropriate parameterizations, this minimum cost path approach is able to extract the intramuscular pathways at subvoxel accuracy, and taking very little time to do so. This makes us more confident that these algorithms are suitable to be incorporated into a CAD tool aiming to support the DIEAP flap preoperative planning task. The detection of the perforators is also a step towards the efficient creation of 3D models which may be of great relevance to the surgical team. In Fig.~\ref{fig:render_3d}, we show a representation of one of the DIEAP trees extracted by the methodology presented in this paper.

\section{Conclusions} \label{sec:conclusion}

In this paper, we proposed a semi-automatic approach for the detection of the DIEAPs requiring minimal user intervention, mainly when compared with the long and tedious manual analysis that technicians and radiologists face. For the best of our knowledge, this is the first CAD dedicated to these small abdominal vessels. We believe that it may reduce the overall time that the characterization of the DIEAPs takes and, additionally, that it supports a more objective and error-free analysis.

After knowing the location of the anterior fascia of the rectus abdominis muscle, we divided the DIEAPs detection problem into two independent challenges: the extraction of the subcutaneous and intramuscular courses. The first was achieved via a centerline tracking procedure employing local analysis of the gradient vectors of vessel enhanced data to find the local vessel direction. Robustness was increased by incorporating a correction framework based on ridge detection in cross sectional images of the vessel. The average Euclidean and Hausdorff distances between the \textit{Ground Truth} annotations and the extracted subcutaneous paths were 0.64 and 1.17 mm, respectively. The extraction of the intramuscular path was addressed by a Frangi Vesselness based minimum cost path approach. Among the considered parameterizations of the cost funtion, the one achieving the best compromise between accuracy and time reached average Euclidean and Hausdorff distances of 0.51 and 1.00 mm, respectively. It took, in average, 3.50 s to perform the task, using an Intel Core i7-4500U CPU @ 1.80 GHz 2.40 GHz with 8 GB of RAM.

Nonetheless, some topics deserve attention. Considering the subcutaneous tracking procedure, attention should be given to the perforators which present a significant course along the fascia. This makes the tracking method unstable at that region due to the corrupted local gradient vectors and it commonly stops earlier than it should, not allowing for a correct retrieval of the coordinates of the region where the perforator leaves the fascia. In the future, we will address caliber and tortuosity estimation after the detection of the DIEAPs, in order to further support the activity of technicians and radiologists involved in the pre-operative planning of DIEAP flaps.

\textbf{Funding}: This work was supported by Funda{\c{c}{\~a}o para a Ci{\^e}ncia e a Tecnologia [Ph.D grant number SFRH/BD/126224/2016].

\textbf{Declaration of interest}: Ricardo J. Ara{\'u}jo and H{\'e}lder P. Oliveira have a patent pending in Europe (EP3352135), United States of America (US2018199997), China (CN108324300), and Japan (JP2018134393).

\section*{References}

\bibliography{comput_med_imag_graph}
\end{document}